\documentclass[12pt]{article}

\usepackage{graphicx}
\usepackage{epsfig}

\begin{document}

\begin{center}
{\Large\bf Description of $\phi$-meson production in hadronic and nuclear collisions
at very high energies} \\
\vspace{0.6cm}

G.H.~Arakelyan$^1$, C.~Merino$^2$, Yu.M. Shabelski$^3$ \\

\vspace{.5cm}
$^1$A.Alikhanyan National Scientific Laboratory \\
(Yerevan Physics Institute)\\
Yerevan, 0036, Armenia\\
e-mail: argev@mail.yerphi.am\\
\vspace{0.1cm}

$^2$Departamento de F\'\i sica de Part\'\i culas, Facultade de F\'\i sica\\
and Instituto Galego de F\'\i sica de Altas Enerx\'\i as (IGFAE)\\
Universidade de Santiago de Compostela\\
15782 Santiago de Compostela\\
Galiza-Spain\\
e-mail: merino@fpaxp1.usc.es \\
\vspace{0.1cm}

$^3$Petersburg Nuclear Physics Institute\\
NCR Kurchatov Institute\\
Gatchina, St.Petersburg 188350, Russia\\
e-mail: shabelsk@thd.pnpi.spb.ru
\vskip 0.9 truecm

{\bf Abstract}
\end{center}

We expose the current experimental and theoretical situation
of the interesting case of the production of $\phi$ mesons
in up to very high energy collisions of hadrons on both nucleon and nuclear targets,
and we present a quantitavely good theoretical description of the corresponding experimental data,
based on the formalism of the well established Quark-Gluon String Model,
that has proved to be valid for a wide energy range.
All the available experimental data for $\phi$-meson production
in hadron-nucleon collisions on the spectra of secondary $\phi$, and
on the ratios of $\phi$/$\pi^-$ and $\phi$/$K^-$ production cross sections,
as well the corresponding ones for $\phi$-meson production on nuclear targets,
are considered. In particular, it is seen that the production of $\phi$-mesons on nuclear targets
presents unusually small shadow corrections for the inclusive density in the central rapidity region.
\vskip -1.5cm

\section{Introduction}
The $\phi$-meson is a system of $s\bar{s}$ quarks, and though
$s$ and $\bar{s}$ quarks have non-zero masses, at the same time their masses are not
large enough to make standard perturbative Quantum Chromodynamics (QCD)
applicable. Thus, one could think of treating the $\phi$-meson as an intermediate case between
soft and hard physics, making it especially interesting to disentangle the open question of the
overlap and transition between the soft (non-perturbative) and the hard (perturbative) regimes.
At the same time, the $\phi$-meson is rarely produced, and it can be sensitive to the production mechanism.

Here we update and complete the theoretical description, already presented in references~\cite{AMShPhi1, AMShPhi2},
of the experimental data on production of vector $phi$-mesons 
a rarely produced system formed of $s$ quark and $\bar{s}$ antiquark, with non-zero masses,
in hadron-nucleon, hadron-nucleus, and nucleus-nucleus collisions for a large energy
region, by including newly published experimental data in the analysis.

For that, we use the
Quark-Gluon String Model (QGSM)~\cite{KTM, K20} formalism, that allows one to make
quantitative predictions on the inclusive densities of different secondaries. In particular, the description
of the experimental data on the production of pseudoscalar mesons $\pi$ and $K$ and baryons $p$,
$\bar{p}$, $\Lambda$, and $\bar{\Lambda}$ was obtained  in~\cite{KaPi, Sh, softpPb, AMPS, AMPSk},
and the same for the production of vector mesons~\cite{aryer, APSh}, and of hyperons,
including the multistrange ones~\cite{ACKS, Sigma}, both in the central and beam fragmentation regions.
Thus, the QGSM provides a consistent description of multiparticle production processes
in hadron-nucleon~\cite{KTM, K20}, hadron-nucleus~\cite{KTMS, Sh1}, and nucleus-nucleus~\cite{Sha, AMPSpl}
collisions up to the currently available high energies of the Relativistic Heavy Ion Collider-RHIC (BNL-USA)
and the Large Hadron Collider-LHC (CERN-Switzerland).

The QGSM analysis gives a quantitatively consistent description of the spectra of secondary $\phi$ mesons,
as well as of the ratios of $\phi$/$\pi^-$ and $\phi$/$K^-$ production cross sections
in $pp$-collisions for a large scope of the initial energy going up to the RHIC and LHC ranges.
In the case of collisions with a nuclear target, the description of the experimental data is also quantitatively
good for both the initial energy dependence and the $A$-dependence of the produced $\phi$-mesons, and,
in particular, a natural explanation appears for the unusually small shadow corrections experimentally
observed in proton-nucleus and nucleus-nucleus collisions at very high energies for the inclusive
density~\cite{softpPb, ACKS, MPSd, MPSpa} (saturation of the inclusive density) in the central rapidity region.

\section{Inclusive meson production in the QGSM formalism}

In order to obtain quantitative results in describing inclusive secondary-hadron spectra integrated
over the secondary-particle transverse momentum, $p_T$, a model for multiparticle production is needed.
It is for that purpose that we have used the QGSM~\cite{KTM, K20} in the numerical calculations presented here.

The QGSM~\cite{KTM, K20, KTMS} is a model based on the Dual Topological Unitarization (DTU), Regge phenomenology, 
and nonperturbative notions of QCD that
succesfully predicts and describes many features of the hadronic processes in a wide energy range. In particular, 
the QGSM allows one to make quantitative predictions on the inclusive densities of different secondaries, both 
in the central and the beam fragmentation regions. 

\begin{figure}[htb]
\vskip -9.5cm
\includegraphics[width=1.0\hsize]{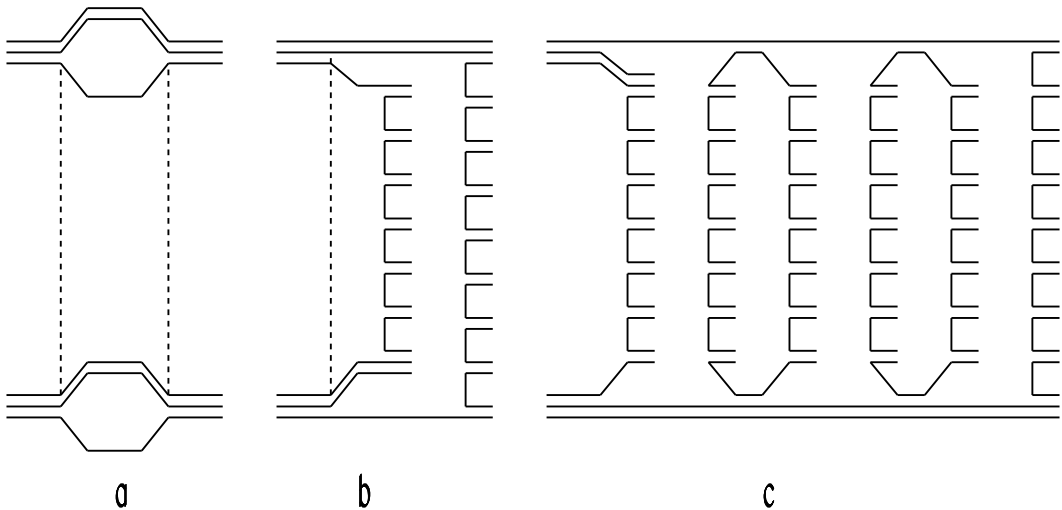}
\vskip -4.0cm
\caption{\label{fig:cil}
(a) Cylinder-type diagram representing a Pomeron exchange within the DTU
classification (the solid lines represent quarks), (b) cut of the 
cylinder-type diagram corresponding to the contribution of one-Pomeron exchange 
to inelastic $pp$ scattering, and (c) the cuts of one of the diagrams for the inelastic 
interaction of the incident proton with the target proton in a $pp$ collision.}
\end{figure}

In the QGSM frame, high energy hadron-nucleon collisions are considered as taking place via the exchange
of one or several Pomerons. Each Pomeron is considered in DTU as a cylinder-type diagram (Fig.~\ref{fig:cil}a),
in which the cylinder boundaries are drawn by the dash-dotted vertical lines.
The surface of the cylinder is schematically depicted by dashed lines, while the solid lines
at the top and bottom of the cylinder represent, respectively, the beam and the target quarks,
which interaction is mediated by the Pomeron exchange.

The cut~\cite{AGK} between Pomerons in a multipomeron diagram results in elastic or diffraction
dissociation processes, while the cut through one (Fig.~\ref{fig:cil}b) or several (Fig.~\ref{fig:cil}c)
Pomerons corresponds to inelastic processes with multiple production of secondaries, the cut of every Pomeron
leading to the production of two showers of secondaries.

For a nucleon target, the inclusive rapidity, $y$, or Feynman-$x$, $x_F$,
spectra of a secondary hadron $h$ have the form~\cite{KTM}:
\begin{equation}
\frac{dn_h}{dy}\ = \
\frac{x_E}{\sigma_{inel}}\cdot \frac{d\sigma}{dx_F}\ = 
\sum_{n=1}^\infty w_n\cdot\phi_n^h (x) \ ,
\label{eq:eq1}
\end{equation}
where $x_E=E/E_{max}$, the functions $\phi_{n}^{h}(x)$ determine the contribution of diagrams
with $n$ cut Pomerons and $w_n$ is the relative weight of these diagrams, which is calculated as
\begin{equation}
w_n=\sigma^n/\sigma_{inel}\ \;,
\label{eq:eq2}
\end{equation}
with $\sigma_{inel}=\sigma_{tot}-\sigma_{el}$.

The cross sections of all inelastic processes corresponding to diagrams where $n \geq 1$ 
Pomerons are cutted can be calculated~\cite{Kar3} with he help of the AGK cutting rules~\cite{AGK}:
\begin{equation}
\sigma^{(n)}_{hN} = \frac{\sigma_{P}}{n \cdot z} \left( 1 - e^{-z} 
\sum_{k = 1}^{n - 1} \frac{z^{k}}{k!} \right)\;. 
\label{eq:eq13}
\end{equation}

The average number of exchanged Pomerons in $pp$ collisions
$\langle n \rangle_{pp}$ slowly increases with the energy.

The parametrization of the Pomeron pole and the calculation of the corresponding elastic and inelastic 
cross sections, are presented in Appendix~A, together with a detailed list of the corresponding references.

The inclusive spectrum of secondaries is then determined by the convolution of diquark, valence quark, 
and sea quark distributions in the incident particles, $u(x,n)$, with the fragmentation functions of 
quarks and diquarks into the secondary hadrons, $G(z)$.
The distribution functions, $u(x,n)$, and the fragmentation functions, $G(z)$, are
determined by the appropriate Reggeon diagrams~\cite{K20, KTMS, Kai}, and
both are constructed using the Reggeon counting rules~\cite{Kai}. 

The diquark and quark distribution functions depend on the number $n$ of cut Pomerons in the
diagrams being considered.
The parametrization of distribution functions of quarks and diquarks in colliding particles 
were obtained in~\cite{AMShPhi1, KTMS}, and they are presented in the Appendix~B.

The parametrizations of quark and diquark fragmentation functions to $\phi$-mesons, together 
with the normalization parameter $a_{\phi}$, were given in \cite{AMShPhi1, aryer} and they are presented 
here in Appendix~C. The value of the parameter $a_{\phi}$ = 0.11 was determined from comparison 
with low energy experimental data, and it has been used in the theoretical calculations for the whole energy region
under consideration.

Note that the quark-antiquark distribution functions $u(x,n)$ differ from the standard $PDF$ 
distributions extracted from a fit to experimental data, since the functions $u(x,n)$ are 
applicable at the rather low $Q^2$ which correspond to soft processes, whereas the $PDF$ 
distributions resulted from describing high-$Q^2$ processes.

For $pp$ collisions, one has
\begin{eqnarray}
\phi_n^{h}(x) &=& f_{qq}^{h}(x_{+},n) \cdot f_{q}^{h}(x_{-},n) +
f_{q}^{h}(x_{+},n) \cdot f_{qq}^{h}(x_{-},n) \nonumber\\
	&+& 2(n-1)\cdot f_{s}^{h}(x_{+},n) \cdot f_{s}^{h}(x_{-},n)\ \ \label{eq:eq3} , \\
x_{\pm} &=& \frac{1}{2}[\sqrt{4m_{T}^{2}/s+x^{2}}\pm{x}]\ \ , 
\label{eq:eq4}
\end{eqnarray}
where $m_T=\sqrt{m^2+p_T^2}$ is the transverse mass of the product hadron
and $f_{qq}$, $f_{q}$, and $f_{s}$ represent the contributions
of, respectively, diquarks, valence quarks and antiquarks, and sea quarks and antiquarks~\cite{K20, Kai}.
In the case of meson-nucleon collisions, the diquark contribution 
$f_{qq}^{h}(x_{+},n)$ in Eq.~(\ref{eq:eq3}) should be replaced by the valence antiquark
contribution $f_{\bar{q}}^{h}(x_{+},n)$.

The functions $f_{qq}$, $f_q$,  $f_{\bar q}$, and $f_{sea}$ are determined
by the convolution of the diquark, quark, and antiquark distribution functions,
$u(x,n)$, with the fragmentation functions to hadron $h$, $G^h(z)$, e.g.,
\begin{equation}
f_i^h(x_+,n)\ =\ \int\limits_{x_+}^1u_i(x_1,n)G_i^h(x_+/x_1) dx_1\ , 
\label{eq:eq5}
\end{equation}
where $i$ stands for diquarks ($qq$), valence and sea quarks ($q$), and valence and
sea antiquarks ($\bar{q}$). The details of the model are presented in~\cite{KTM, K20, KaPi, Sh, ACKS}.

To describe the nucleon interaction with nuclear targets, use is made of Gribov-Glauber multiple-scattering
theory. It permits considering collisions with nuclei as a superposition of interactions involving
various numbers of target nucleons~\cite{Sh3, BT, Weis, Jar}.

In the case of nucleus-nucleus collisions, this approach makes it possible to consider this interaction
as a superposition of individual nucleon-nucleon interactions. Although an analytic summation of all
diagrams is impossible in this case~\cite{BraunSh}, the most important classes of diagrams can be
summed analytically in the so-called rigid nucleus approximation~\cite{Alk}, 
that we use here.

In calculating inclusive spectra of secondary particles produced
in $pA$ collisions, we should consider the possibility of cutting
one or several Pomerons in each of the $\nu$ blocks of the inelastic proton-nucleon
interaction.

\begin{figure}[htb]
\vskip -9.0cm
\includegraphics[width=1.0\hsize]{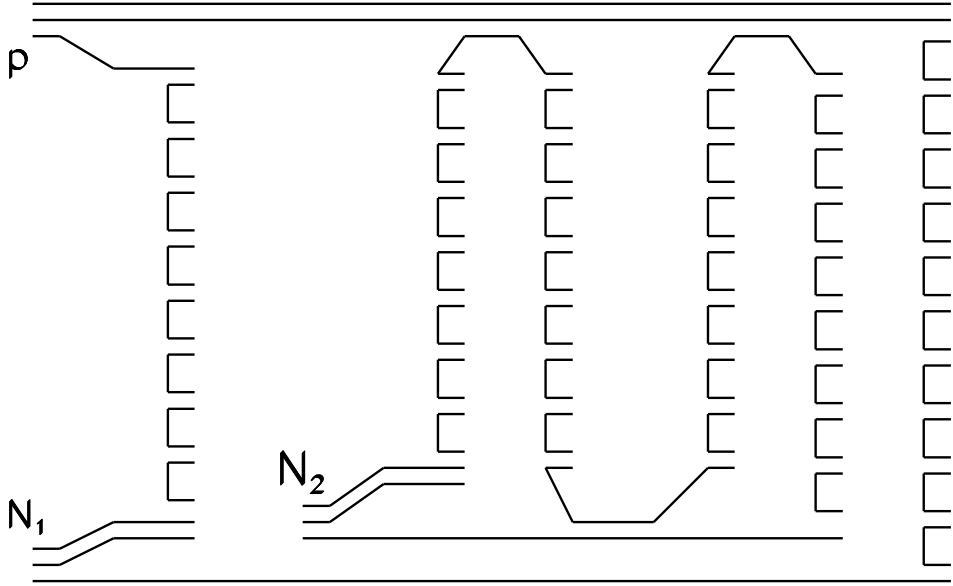}
\vskip -2.5cm
\caption{\label{fig:fig2} One of the diagrams corresponding to the inelastic
interaction of one incident nucleon with two target nucleons $N_1$ and $N_2$
in a $pA$ collision.}
\end{figure}

For example, Fig.~\ref{fig:fig2} shows one of the diagrams that contribute
to the inelastic interaction of a beam proton with two target nucleons.
One Pomeron is cut in the proton-nucleon$_1$ interaction block,
and two Pomerons are cut in the proton-nucleon$_2$ interaction block. 

The process shown in Fig.~\ref{fig:fig2} satisfies the condition that the absoptive
part of the hadron-nucleus amplitude is determined by the combination of
the absorptive parts of the hadron-nucleon amplitude~\cite{Sh3, BT, Weis, Jar}.

The contribution of the diagram in Fig.~\ref{fig:fig2} to the inclusive cross sections 
has the form
\begin{eqnarray}
\frac{x_E}{\sigma_{prod}^{pA}}\cdot\frac{d \sigma}{dx_F} & = & 2\cdot
W_{pA}(2)\cdot w^{pN_1}_1\cdot w^{pN_2}_2\cdot\left\{
f^h_{qq}(x_+,3)\cdot f^h_q(x_-,1)\right. + \nonumber\\ \nonumber & + &
f^h_q(x_+,3)\cdot f^h_{qq}(x_-,1) + f^h_s(x_+,3)\cdot[f^h_{qq}(x_-,2) +
f^h_q(x_-,2) + \\ & + & 2\cdot f^h_s(x_-,2)] \left. \right\} \;,
\end{eqnarray}
where $W_{pA}(2)$ is the probability for the interaction of the proton beam
with two target nucleons.

According to multiple-scattering theory, one has:
\begin{equation}
W_{pA}(\nu) = \sigma^{(\nu)}/\sigma_{prod}^{pA} \;
\end{equation}
(see~\cite{softpPb} for numerical examples). Here,
\label{eq:eq15}
\begin{equation}
\sigma^{(\nu)} = \frac1{\nu !} \int d^2b\cdot [\sigma^{pN}_{inel}\cdot T(b)]^{\nu}\cdot
e^{-\sigma^{pN}_{inel}\cdot T(b)} \ \;,
\label{eq:eq16}
\end{equation}
while
$\sigma_{prod}^{pA}$ is the cross section for the production of secondary particles
in $pA$ collisions.

It is important to take into account all diagrams featuring all possible
configurations of Pomerons and their respective permutations.
The quark and diquark distributions and fragmentation functions are identical to those
in the case of $pN$ interaction.

The growth of the total number of interacting Pomerons is given by
\begin{equation}
\langle n \rangle_{pA} \sim
\langle \nu \rangle_{pA} \cdot \langle n \rangle_{pN} \;,
\label{eq:eq17}
\end{equation}
where $\langle \nu \rangle_{pA}$ is the average number of inelastic collisions within
the target nucleus (about four for heavy nuclei at fixed target energies in the range $10^2-10^3 GeV$).

In the case of nucleus-nucleus scattering in the projectile-fragmentation region,
we use the approach~\cite{Sha, Shab, JDDS} in which a beam of
independent nucleons interacts with the target nucleus,
this corresponding to the rigid-target approximation of
Glauber's theory~\cite{Alk}.

In the target fragmentation region, the beam of independent target nucleons
interacts with a projectile nucleus. Thus, these two contributions coincide
in the midrapidity region. 

If the initial energy is not very high, corrections associated with the
energy-conservation law play an important role.
This approach was succesfully used in~\cite{JDDS} to describe
$\pi^{\pm}$, $K^{\pm}$, $p$, and $\bar{p}$ production in $PbPb$
collisions at 158 GeV/c per nucleon.

In the case of $\phi$-meson production in the midrapidity region,
we disregard the contributions of diffractive-dissociation processes.

\section{Hadroproduction of $\phi$-mesons on proton target}
\label{sec:sec3}

In this section we compare the QGSM calculations with the experimental data on 
$\phi$ inclusive  cross sections in $\pi p$ and $pp$ collisions at different 
energies up to the LHC range.
\begin{figure}[htb]
\label{pi140}
\vskip -9.0cm
\hskip 3.25cm
\includegraphics[width=0.775\hsize]{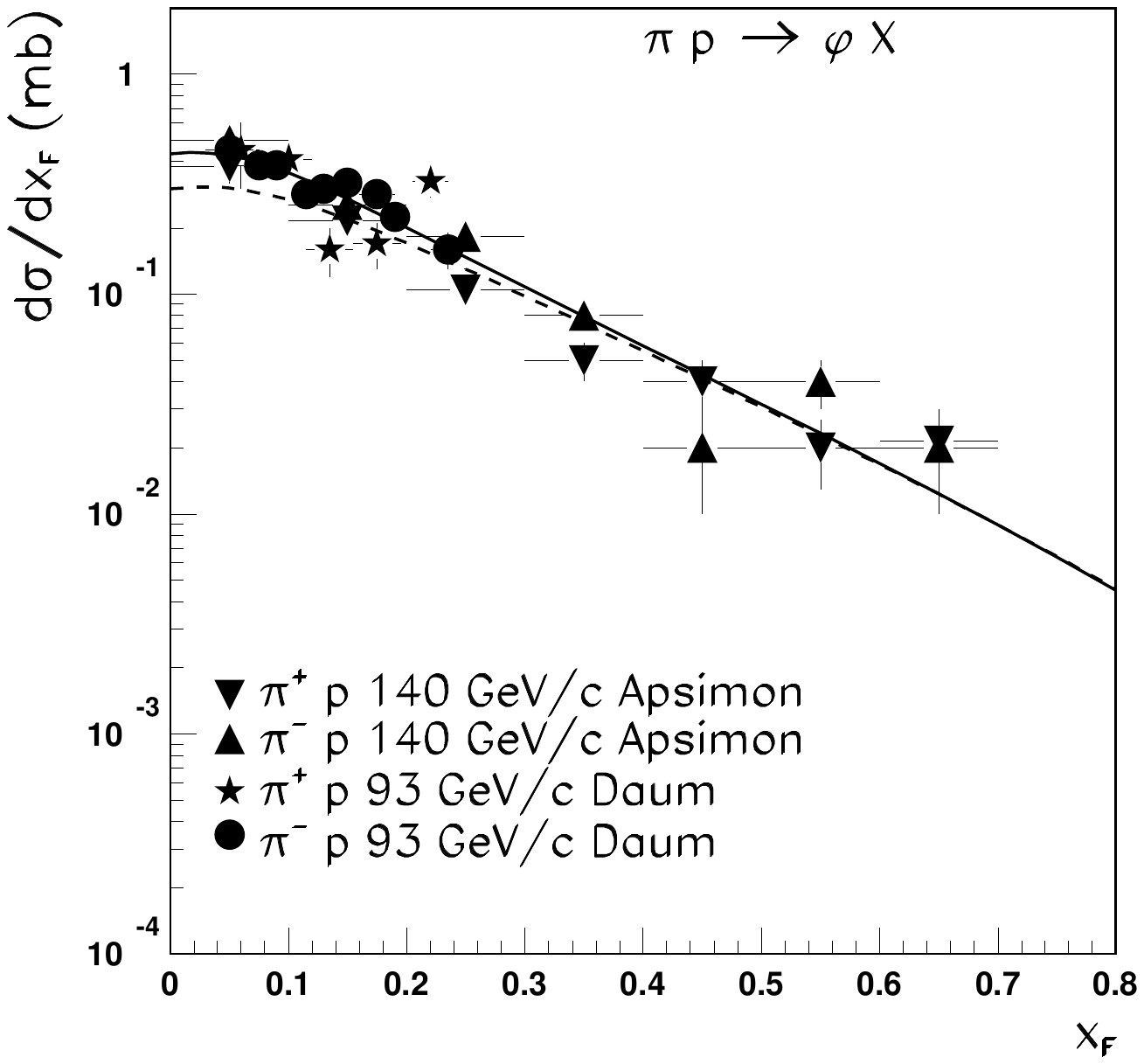}
\vskip -8.5cm
\hskip 3.25cm
\includegraphics[width=0.775\hsize]{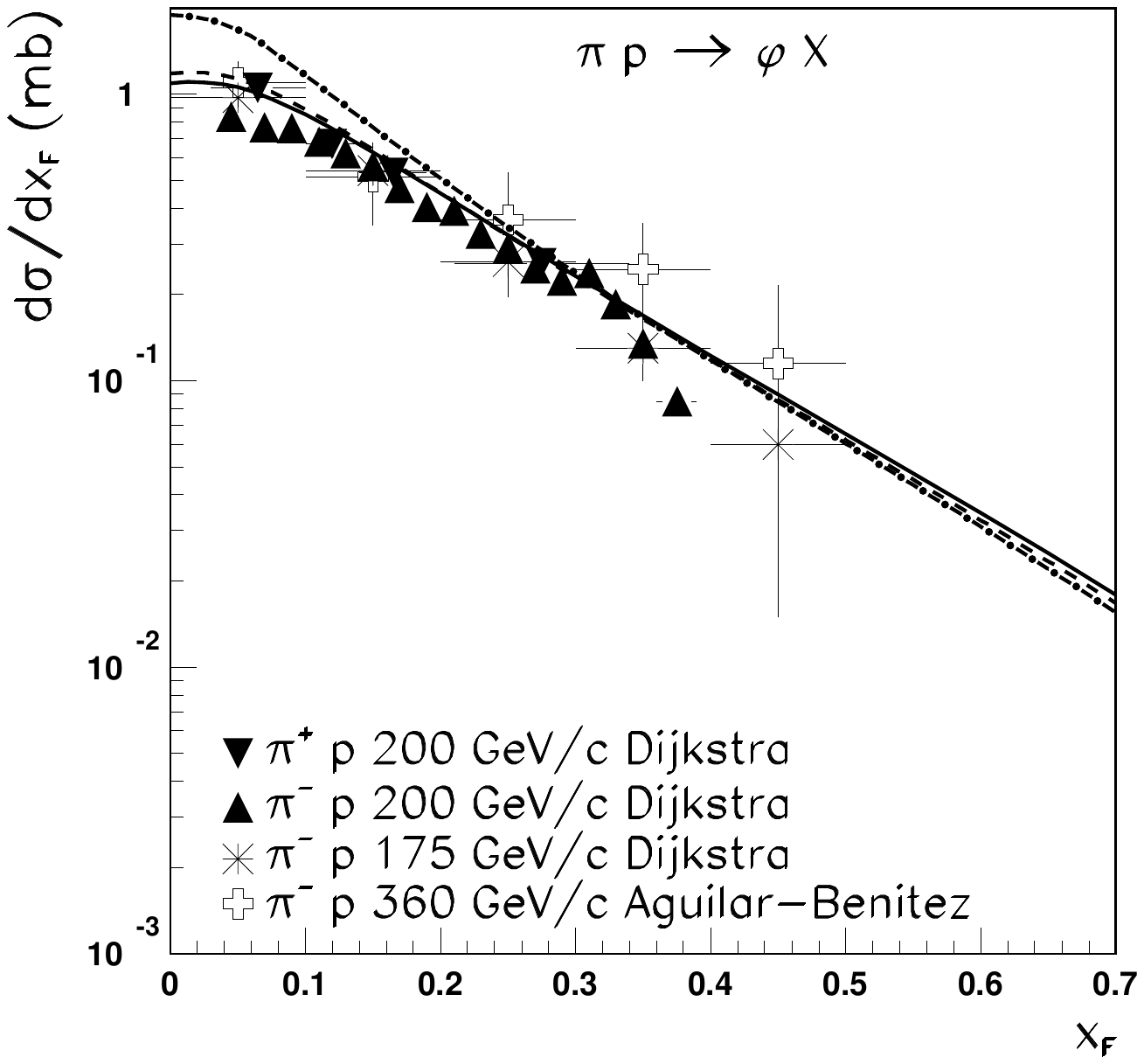}
\vskip -0.75cm
\caption{\label{fig:pi140} Different experimental data on  
$x_F$-spectra of $\phi$-mesons produced in $\pi^{\pm} p$ collisions at
different energies~\cite{daum, apsimon, dijkstra}, compared to the results
corresponding calculations based on the QGSM (see the main text for the description of
the different experimental data sets and theoretical curves).}
\end{figure}

In the upper panel of Fig.~\ref{fig:pi140} we present the experimental data for the differential cross section 
$d\sigma/dx_F$ of $\phi$-mesons produced in $\pi^{\pm} p$ collisions 
at initial momenta of $\pi$-mesons 93~\cite{daum}, 
and 140~GeV/c~\cite{apsimon}. 
The results of the corresponding QGSM calculations 
at pion beam momenta 93 and 140~GeV/c are shown by dashed and solid curves,
respectively. 
The theoretical curves 
are in good agreement with experimental data,
both for 93 and 140 GeV/c. 

In the lower panel of Fig.~\ref{fig:pi140}, the data on $\phi$-meson production at the higher
pion beam initial momenta of 175, 200, and 360 GeV/c are also presented. We see that the experimental data for these
three pion beam momenta are in agreement with the results of the theoretical calculations based on the QGSM
shown by solid, dashed, and dashed-dotted curves, respectively.
Here, the theoretical curves are in general agreement with the data, except for the small
$x_F$ region, where we predict a significant increase of $d\sigma/dx_F$ from 200 to 360 GeV/c that
is not seen in the experimental data.
\begin{figure}[htb]
\vskip -11.0cm
\hskip 2.0cm
\includegraphics[width=1.0\hsize]{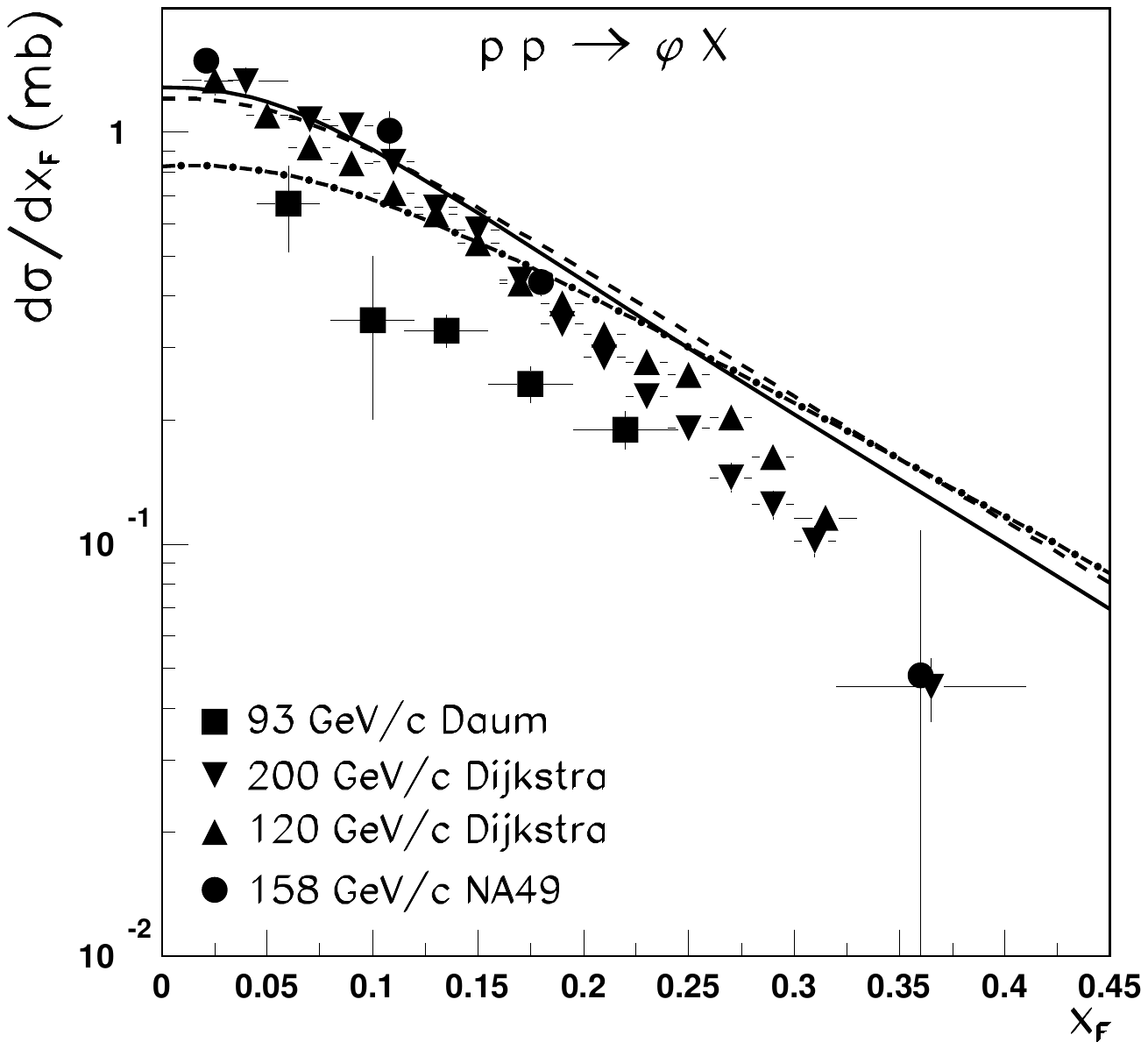}
\vskip -1.0cm
\caption{\label{fig:phi158}
The experimental data~\cite{daum, dijkstra, afanas, anticic} on the $x_F$ spectra
of $\phi$-mesons produced in $pp$ collisions at 93, 120, 158, and 200 GeV/c,
compared to the results of corresponding calculations based on the QGSM at 93, 158, and 200 GeV/c.}
\end{figure}

The QGSM description of the experimental data on the $x_F$ dependence of
$d\sigma/dx_F$-spectra of $\phi$-mesons in $pp$-collision measured at 93~\cite{daum},
120 and 200 GeV/c~\cite{dijkstra}, and 158 GeV/c~\cite{afanas, anticic}, 
is presented in Fig.~\ref{fig:phi158}. The solid 
curve corresponds to the proton beam momenta 93 GeV/c, the dashed curve to 158 GeV/c, 
and the dashed-dotted curve to 200 GV/c. The agreement of the theoretical
calculation with the data is reasonable at small $x_F$, at large $x_F$ the experimental
data falling down faster than the theory, except for the data at 93 GeV/c,
when even at small $x_F$ the agreement is not good. One can appreciate
some contradiction between the data on the production of $\phi$-mesons in $pp$ (Fig.~\ref{fig:phi158}) and
those in $\pi p$ (upper panel of Fig.~\ref{fig:pi140}) collisions, obtained both by
the same experiment~\cite{daum}. The results of the calculations based on the QGSM at the NA49 Collaboration energy
are in agreement with the Monte Carlo predictions obtained by using the LUCIAE model~\cite{luciae}. 
\begin{figure}[htb]
\vskip -11.0cm
\hskip 2.0cm
\includegraphics[width=1.0\hsize]{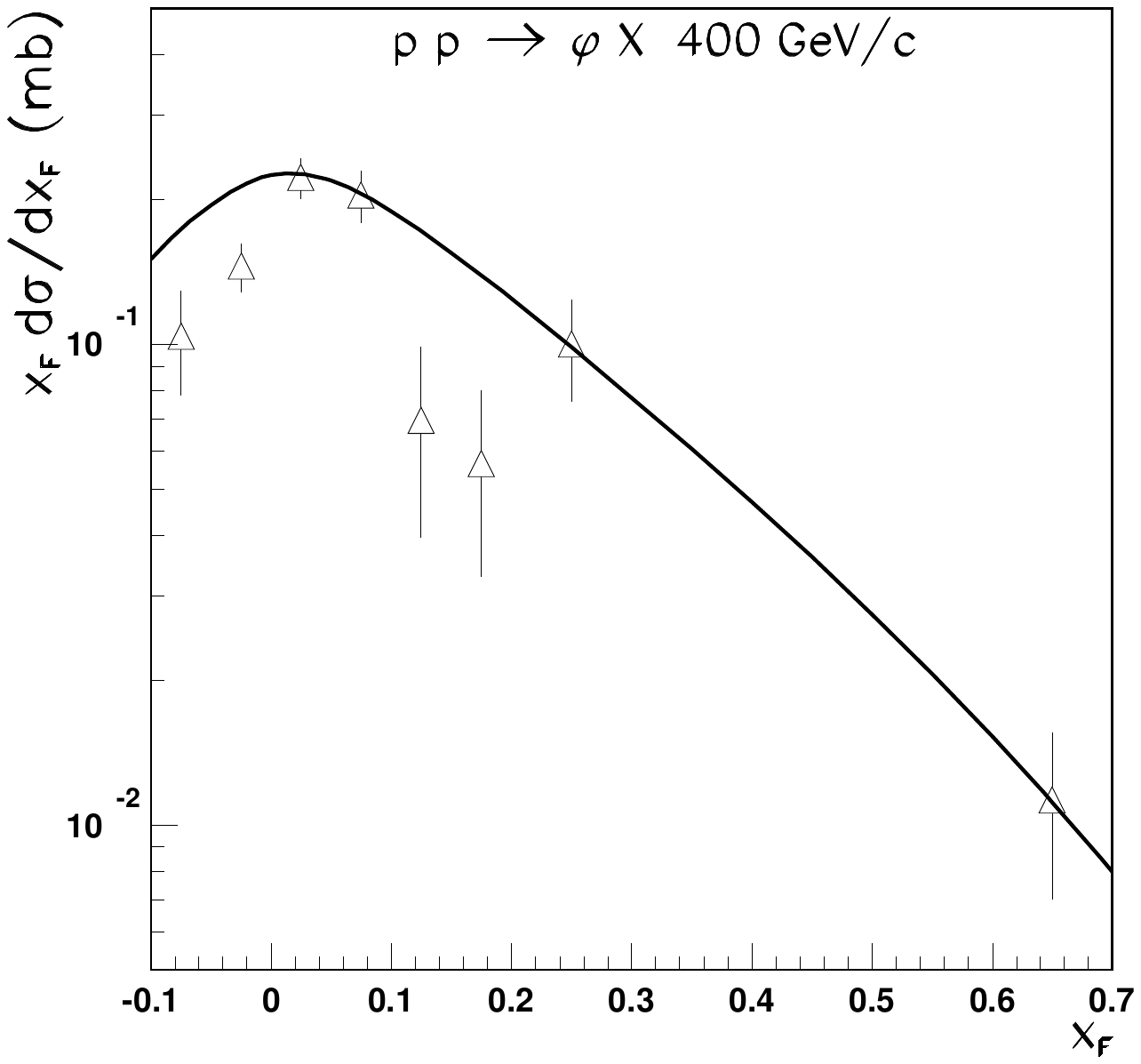}
\vskip -1.0cm
\caption{\label{fig:phi400}
Experimental data~\cite{afanas, anticic}
on the $x_F$-spectra of $\phi$-mesons 
produced in $pp$ collisions at a proton beam momentum 400 GeV/c compared to
the results of the corresponding QGSM calculation.}
\end{figure}

In Fig.~\ref{fig:phi400}, we compare the QGSM calculations to the experimental data on the inclusive spectra  
$x_F\cdot d\sigma /dx_F$ of $\phi$-mesons in $pp$-collisions at 400 Gev/c~\cite{afanas, anticic}
The agreement of the theoretical curve 
with the data is good.

\begin{figure}[htb]
\vskip -11.0cm
\hskip 2.0cm
\includegraphics[width=1.0\hsize]{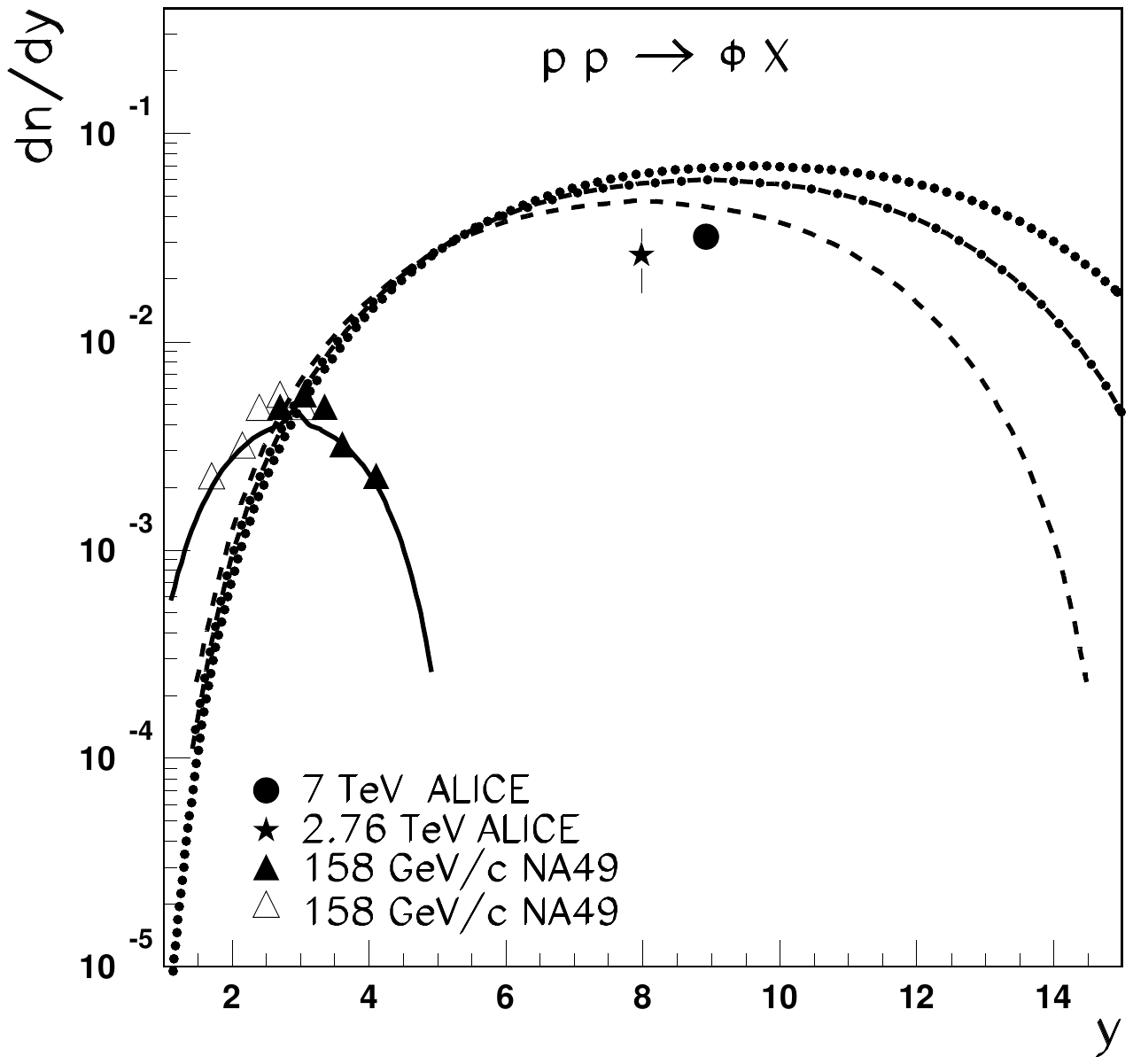}
\vskip -1.0cm
\caption{\label{fig:dndy7TeV}
Experimental data~\cite{afanas, anticic}
on the $y$-spectra $dn/dy$ of $\phi$-mesons
produced in $pp$ collisions at 158 GeV/c, and the results of the corresponding 
calculation based on the QGSM (solid line). The dashed and dashed-dotted lines 
represent the results of the QGSM calculations at 2.76 and 7~Tev, and they are compared with 
corresponding experimental points in the central rapidity region,
$|y|\le$ 0.5 \cite{alpppb, alfik}. The dotted line is the QGSM prediction for 14~TeV.}
\end{figure}

In Fig.~\ref{fig:dndy7TeV}, the rapidity dependence of the density $dn/dy$ of $\phi$-mesons 
produced in $pp$-collisions at 158 GeV/c~\cite{afanas, anticic} is compared to the results 
of the corresponding calculation based on QGSM (solid curve). The agreement is quite reasonable. 

We also compare two new experimental points by ALICE Collaboration on the 
density $dn/dy$ of $\phi$-mesons produced in $pp$-collisions at 2,76~\cite{alpppb} and  
7~TeV~\cite{alfik}, with the results of the corresponding QGSM calculations. The dashed curve represents
the theoretical results for 2.76~TeV, and the dashed-dotted curve those for 7~TeV.
The dotted curve shows the model prediction for 14~TeV.

As it can be seen in Fig.~\ref{fig:dndy7TeV}, the results of the QGSM calculations are remarkably higher than
the experimental data, which were measured for $p_T \geq $ 0.4~GeV/c, and extrapolated to $p_T$ = 0~\cite{alpppb, alfik}. 
  
Generally, the QGSM description of the experimental data in the 
considered energy region shown in Figs.~\ref{fig:pi140}$-$\ref{fig:dndy7TeV} is consistent.
The description of the large $x_F$ region is rather good in the case of $\pi p$ collisions,
while for $pp$ collisions there is not experimental point at $x_F>$ 0.35,
except for one point at 400 GeV/c that is in agreement with the QGSM curve.

The QGSM predictions have always been rather reliable for describing
$\pi$-meson production up to the LHC energies~\cite{KaPi, Sh, MPS}. Thus we have used the
published results of the QGSM calculations on $\pi$-meson yields to compute the ratios $\phi$/$\pi$. 

\begin{figure}[htb]
\vskip -7.0cm
\hskip 3.0cm
\includegraphics[width=0.665\hsize]{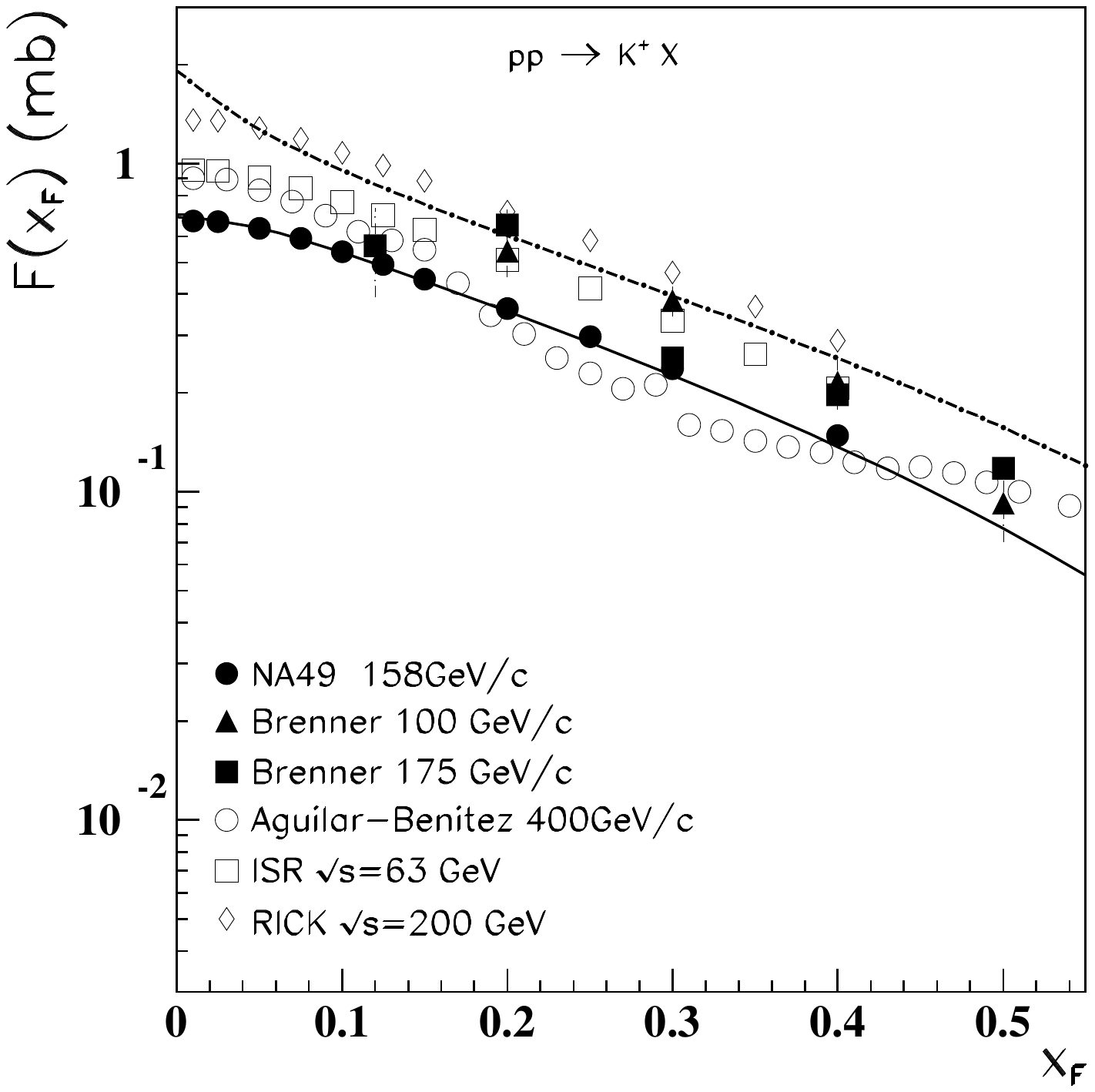}
\vskip -6.25cm
\hskip 3.0cm
\includegraphics[width=0.665\hsize]{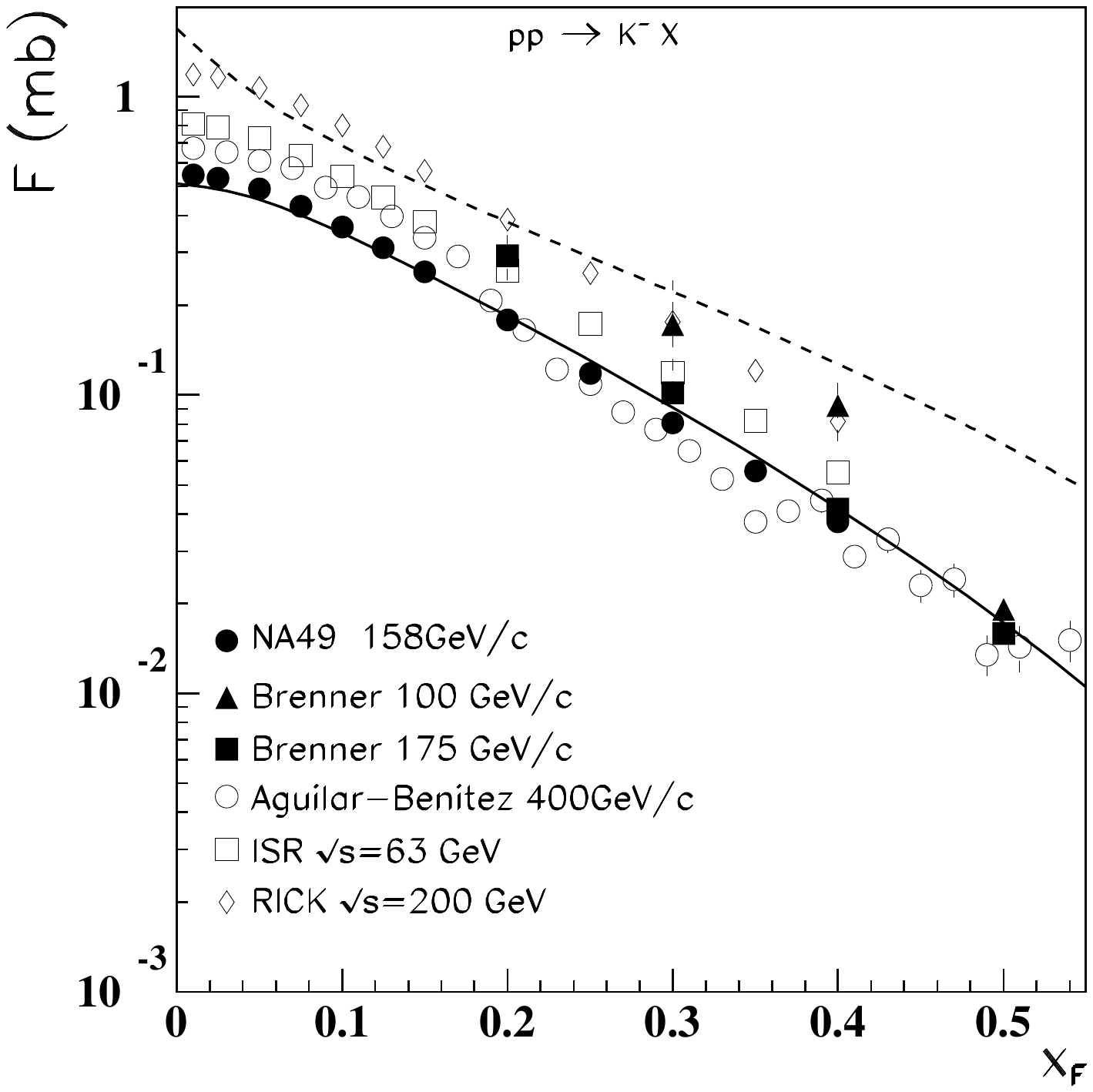}
\vskip -1.25cm
\caption{\label{fig:K}
Experimental data on the invariant cross section of $K^+$ (upper panel) and $K^-$ (lower panel) 
mesons produced in $pp$ collisions at different energies~\cite{afanas, anticic, brenner}
compared to the results of the corresponding calculations based on the QGSM.
Solid lines correspond to the QGSM result at 158 GeV/c ($\sqrt{s}$ = 17.3~GeV),
while dashed lines correspond to the QGSM calculation at RHIC energy ($\sqrt{s}$ = 200 Gev).}
\end{figure}

The spectra of kaons are presented in Fig.~\ref{fig:K}. We present the QGSM description
of the experimental data on the inclusive spectra of $K^+$ (upper panel) and $K^-$
(lower panel) mesons in $pp$-collisions on a wide energy range from $\sqrt{s}$ = 100 up to 
200~GeV~\cite{afanas, anticic, brenner}. 
The integrated over $p_T$ RHIC data at $\sqrt{s}$ = 200~GeV have been taken from~\cite{afanas, anticic}. 
\begin{figure}[htb]
\vskip -4.75cm
\hskip 3.0cm
\includegraphics[width=0.58\hsize]{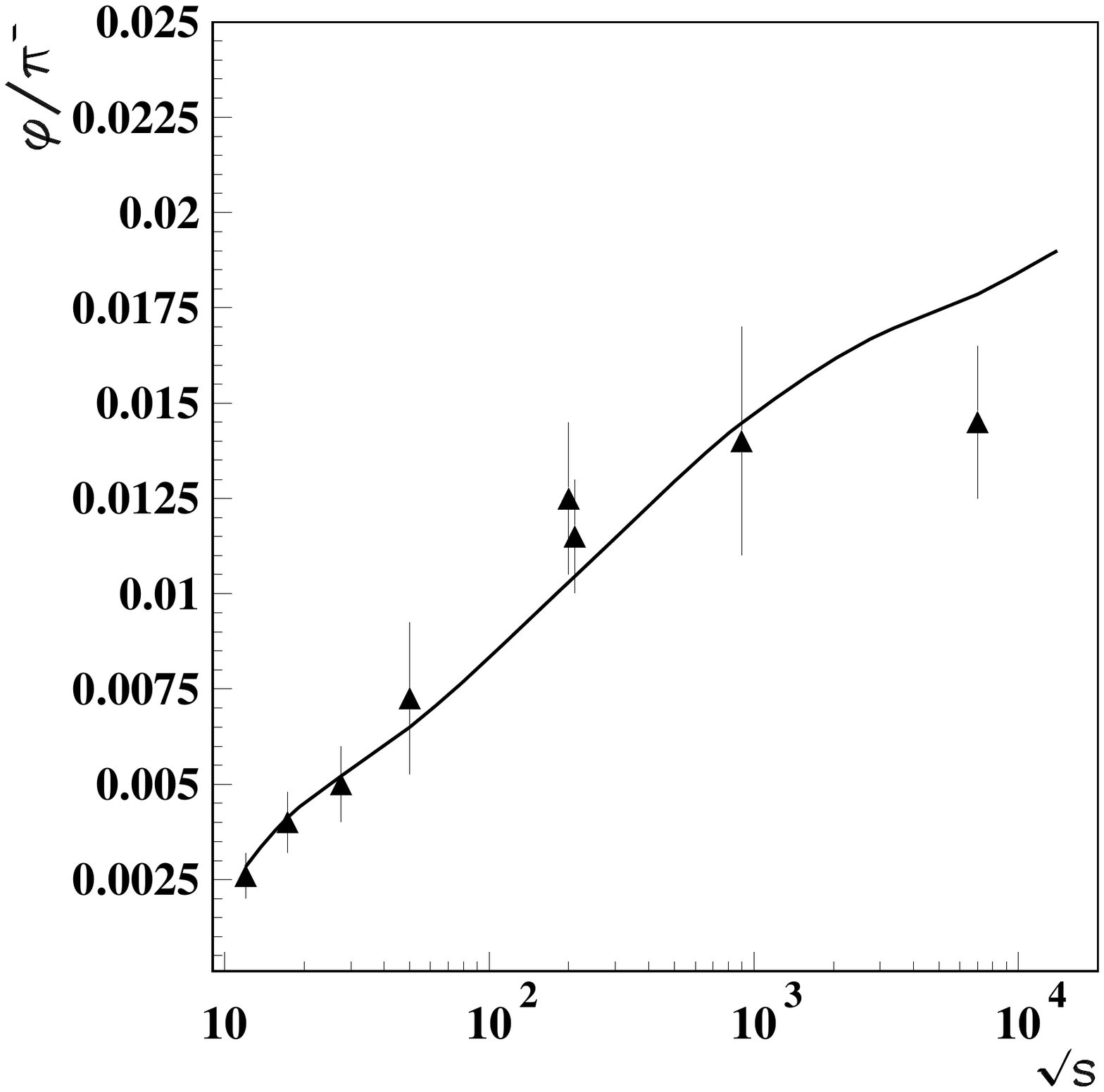}
\vskip -4.5cm
\hskip 3.0cm
\includegraphics[width=0.58\hsize]{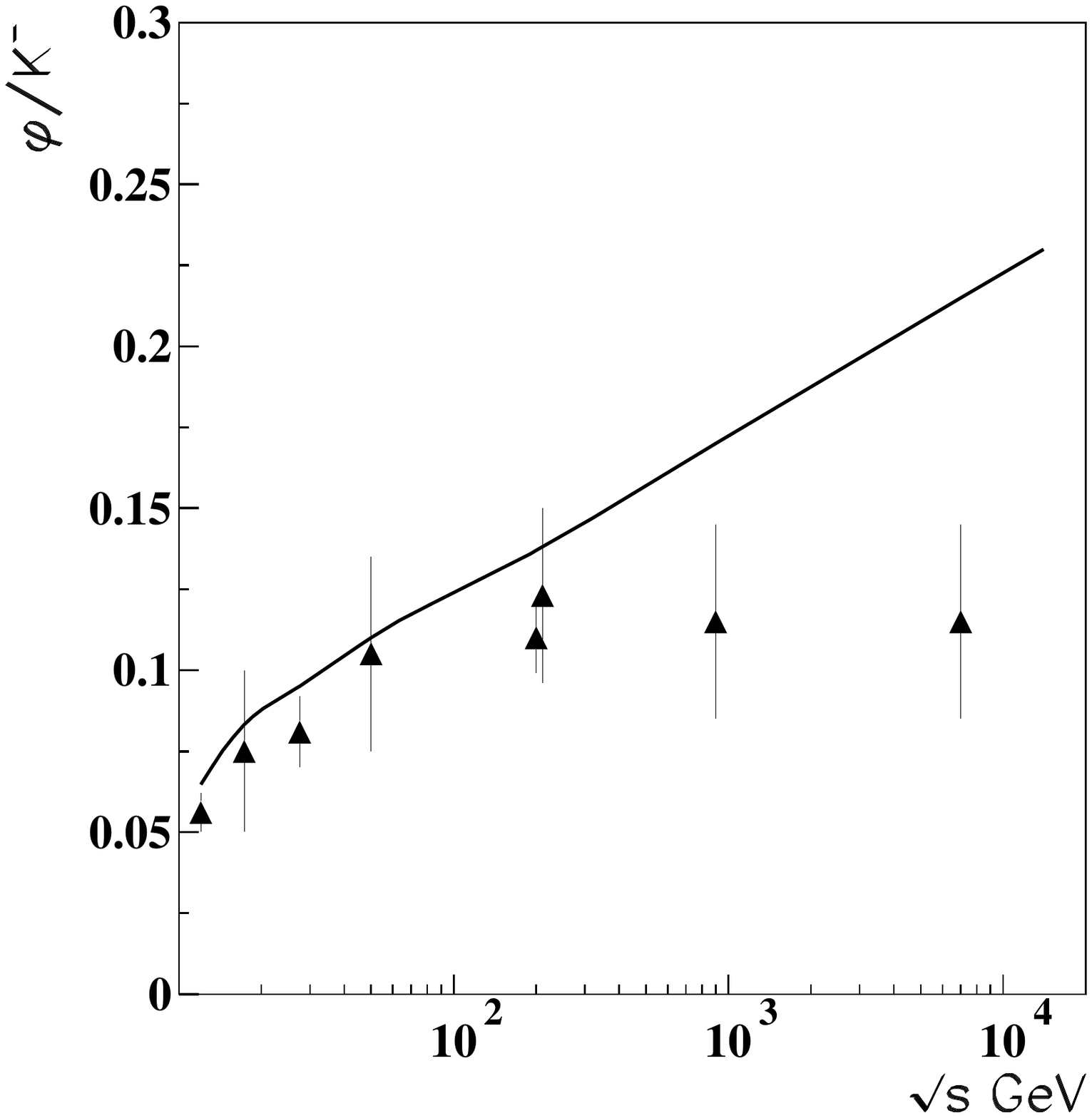}
\vskip -0.75cm
\caption{\label{fig:ratios}
Experimental data~\cite{alfik, starphi, alvec} on the the $\sqrt{s}$ dependence
of $\phi/\pi^-$ (upper panel) and $\phi/K^-$ (lower panel) cross section ratios in $pp$ collisions,
compared to the results of the corresponding calculations based on QGSM.}
\end{figure}

We present the results of the QGSM calculations at the two different proton beam momenta (energies): 
158 GeV/c ($\sqrt{s}$ = 17.3~GeV), by solid curves, and RHIC energy ($\sqrt{s}$ = 200~Gev),
by dashed curves. For both $K^+$ and $K^-$ spectra, the QGSM curves are in reasonable agreement 
with the data. One has to note that some disagreement between the normalisations
of the experimental data by the NA49 Collaboration~\cite{afanas, anticic}
and of the data at 100 and 175 GeV/c~\cite{brenner} exists.
Generally, the experimental spectra of both $K^+$ and $K^-$ increase with the initial beam momentum (energy), 
what is in agreement with the results of our calculations.

In the QGSM the calculated density dn/dy is integrated over $p_T$, so the direct comparison of the results 
of the calculations based on QGSM to the data would be inconsistent. 

\begin{table}
\caption{\label{Table1} Experimental data on the inelastic density $dn/dy$ for $\phi$-meson
production by the ALICE Collaboration at $\sqrt{s}$ = 2.76~\cite{alpppb} and 7~Tev~\cite{alfik},
compared with the results of the corresponding QGSM calculations.}
\begin{center}
\begin{tabular}{llll}
\hline
Reactions & Energy & Experimental data & QGSM \\
	&  $\sqrt{s}$, $TeV$ & on $dn/dy$ ($|y|\leq 0.5$)  &  \\
\hline
p + p &  2.76 & 0.0260 $\pm$ 0.0004 $\pm$ 0.003~\cite{alpppb} & 0.032 \\ 

p + p &  7.00 & 0.032 $\pm$ 0.0004 $\pm$ 0.004~\cite{alfik}  & 0.05 \\ 
\hline
\end{tabular}
\end{center}
\end{table}

In Table~1, the experimental points for the density $dn/dy$ of $\phi$-meson inelastic production by the 
ALICE Collaboration at  $\sqrt{s}$ = 2.76~\cite{alpppb} and 7~Tev~\cite{alfik}, are compared with the results
of the corresponding QGSM calculations. It can be seen that the QGSM results are rather higher than the
experimental points. This is quite unexpected, since the QGSM calculations show in general a quite good agreement
with the experimental data for proton-nucleus and nucleus-nucleus collisions both at RHIC and LHC energies
(see next sections). May be, this is connected with the extrapolation to $p_T$ = 0 of the experimental data
actually measured at transverse momenta $p_T \geq 0.4$ GeV/c. The experimental data are shown in Table~1
as they are presented, after extrapolation, in the original experimental papers.

The energy dependence of the production cross section ratios of $\phi/\pi^-$ (upper panel)~\cite{starphi, alvec} and 
$\phi/K^-$ (lower panel)~\cite{alfik, starphi} in $pp$-collisions are presented in Fig.~\ref{fig:ratios}, where 
the corresponding QGSM description is shown by solid curves. The energy dependences of our curves for $\phi/\pi^-$ 
and $\phi/K^-$ are similar, since the ratio of $K/\pi$ production depends rather weakly on the initial beam momentum.

One can see some disagreements of the QGSM curves with the data at high LHC energies. The 
ratios for $\phi/\pi^-$ and $\phi/K^-$ predicted by the QGSM increase from $\sqrt{s}$ = 0.9 to 7~TeV,
whereas for the experimental values the ratios are practically the same and do not grow 
with energy in the LHC energy range. 

These discrepancies can also be connected with the influence of the kinematical boundaries
at relatively low $p_T$, and they should be, at least in part, smaller when we consider the ratios
$\phi/\pi^-$ or $\phi/K^-$ with the same kinematical restrictions.

\section{The $\phi$-meson production in hadron collisions with nuclear targets up to RHIC energies}
\label{sec:sec4}

We analyze experimental data by the HERA-B Collaboration on $\phi$-meson production in collisions of protons 
with $C$, $Ti$, and $W$ nuclei. These data were integrated over the whole range
of the transverse momentum $p_T$ at $\sqrt{s}=41.6$~GeV~\cite{HERAb}.

\begin{figure}[htb]
\vskip -5.75cm
\includegraphics[width=1.0\hsize]{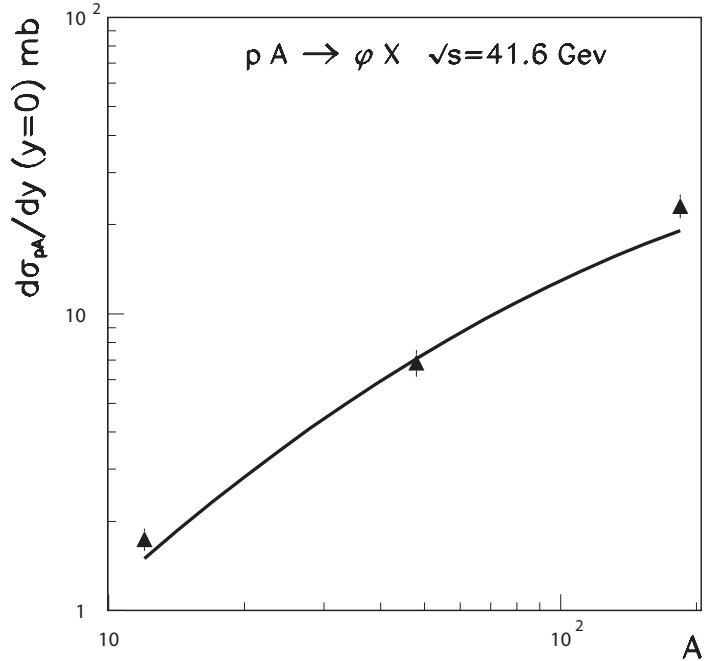}
\vskip -6.0cm
\caption{\label{fig:Adependence}
Experimental data on the $A$-dependence of the cross section for $\phi$-meson production,
$d\sigma_{pA}/dy$ $(y \simeq 0)$, at the enrgy $\sqrt{s}$ = 41.6 GeV~\cite{HERAb} (triangles)
and results of corresponding calculation based on the QGSM (curve).}
\end{figure}

The $A$ dependence of the cross section $d\sigma_{pA}/dy (y \simeq 0)$ 
is shown in Fig.~\ref{fig:Adependence} along with the results of corresponding calculations based on QGSM.
These results are also presented in Table~2.
\vskip 0.75cm

\begin{table}
\caption{\label{Table2} The experimental data of the HERA-B Collaboration~\cite{HERAb}
on $\phi$-meson production in $pA$ collisions at $\sqrt{s}=41.6$ GeV
and the results of corresponding calculations based on the QGSM.}
\begin{center}
\begin{tabular}{lll}
\hline
Reaction & Experimental data $d\sigma_{pA}/dy$ & QGSM \\
	&  $(|y|\leq 0.5), mb$ &  \\
\hline
p + C & 1.74 $\pm$ 0.15  & 1.5 \\

p + Ti & 6.85 $\pm$ 0.7 & 7.1 \\

p + W & 23.5 $\pm$ 2.1 & 19.1  \\
\hline
\end{tabular}
\end{center}
\end{table}

\begin{figure}[htb]
\vskip -10.5cm
\hskip 2.75cm
\includegraphics[width=1.0\hsize]{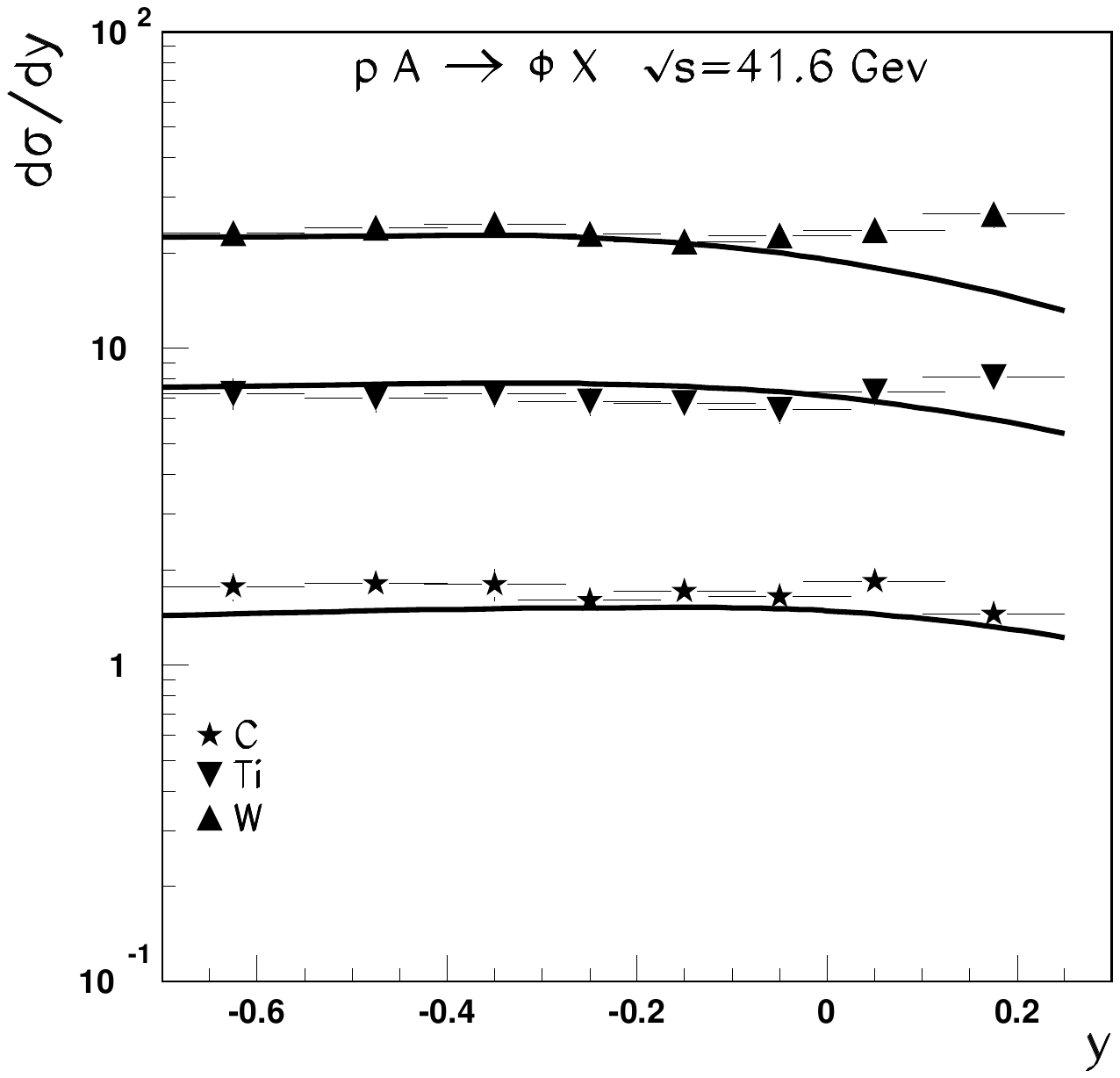}
\vskip -1.0cm
\caption{\label{fig:pAphi41g}
Experimental data on the $y$ dependence of the cross section for $\phi$-meson production,
$d\sigma/dy$, in $pA$ collisions on different nuclei ($A= C$, $Ti$, $W$),
at the energy $\sqrt{s}$ = 41.6 GeV~\cite{HERAb} (points) and results of 
corresponding calculations based on QGSM (curves).}
\end{figure}

In Fig.~\ref{fig:pAphi41g}, experimental data on the rapidity dependence of the inclusive cross section
for $\phi$-meson production, $d\sigma/dy$, in $pA$ collisions for different $C$, $Ti$,
and $W$ targets, and for rather small rapidity ranges, are given along with the results
of corresponding calculations based on the QGSM. 
The theoretical results compare resonably with these experimental data everywhere, 
with the exception of some differences in the positive region of $y\geq 0$ that slitghly increase
with the atomic number $A$ of the target, and that become more apparent for the $W$ nucleus.

In the case of the production of particles such as pions and kaons, which make a dominant
contribution to the mean multiplicity, a new shadowing effect, explained by A.~Capella, A.~Kaidalov, and 
J.~Tran~Thanh~Van in ref.~\cite{CKTr}, appears starting from an energy of $\sqrt{s}= 40$
to $60\ GeV$.

In the case of $\phi$-meson production this new
screening effect is not noticeable, even in the region of the RHIC energies,
but it appears at the LHC energies, which will be discussed in detail
in the next section. 

Let us now consider the experimental data on $\phi$-meson production in heavy ion collisions
at energies from $\sqrt{s}$ = 17.3 to 200~GeV.

In Table~3, the experimental values of the midrapidity inclusive densities,
$dn/dy$ $(|y|\leq 0.5)$, for $\phi$-meson production in central nucleus-nucleus collisions,
obtained by the NA49 Collaboration ($\sqrt{s}$ = 17.3 GeV~\cite{afanas, anticic}),
and those obtained at RHIC (STAR and PHENIX Collaborations, $\sqrt{s}$ = 62.4~GeV~\cite{star2009},
130~GeV~\cite{star130, star130adl}, and 
200~GeV~\cite{starphi, star2009}), are compared with the results of corresponding calculations.
based on the QGSM.

\begin{table}
\caption{\label{Table3} Experimental values of $dn/dy$, $|y|\leq 0.5$, 
for $\phi$-meson production in central nucleus-nucleus collisions 
at different energies, and results of corresponding calculations based on the QGSM.}
\begin{center}
\begin{tabular}{lllll}
\hline
Reaction & Centrality & Energy & Experimental data & QGSM \\
	& \hspace{0.5cm} $\%$ &  $\sqrt{s}, GeV$ &  on $dn/dy$ $(|y|\leq 0.5)$ & \\
\hline
Pb + Pb & 0$-$5\%  & 17.3 & 2.35 $\pm$ 0.15, \cite{afanas, anticic}  & 2.764 \\

Au + Au & 0$-$20\% & 62.4 & 3.52 $\pm$ 0.08 $\pm$ 0.45, \cite{star2009}  & 3.36 \\

Au + Au & 0$-$11\% & 130. & 5.73 $\pm$ 0.37, $\pm$ 0.57, \cite{star130, star130adl}  &6.15 \\

Au + Au & 0$-$5\% & 200. & 7.95 $\pm$ 0.11 $\pm$ 0.73, \cite{star2009} & 8.12 \\

Au + Au & 0$-$5\% & 200. & 7.70 $\pm$ 0.30, \cite{starphi} & 8.12  \\
\hline
\end{tabular}
\end{center}
\end{table}

\begin{figure}[htb]
\label{phi158ys}
\vskip -6.25cm
\includegraphics[width=1.0\hsize]{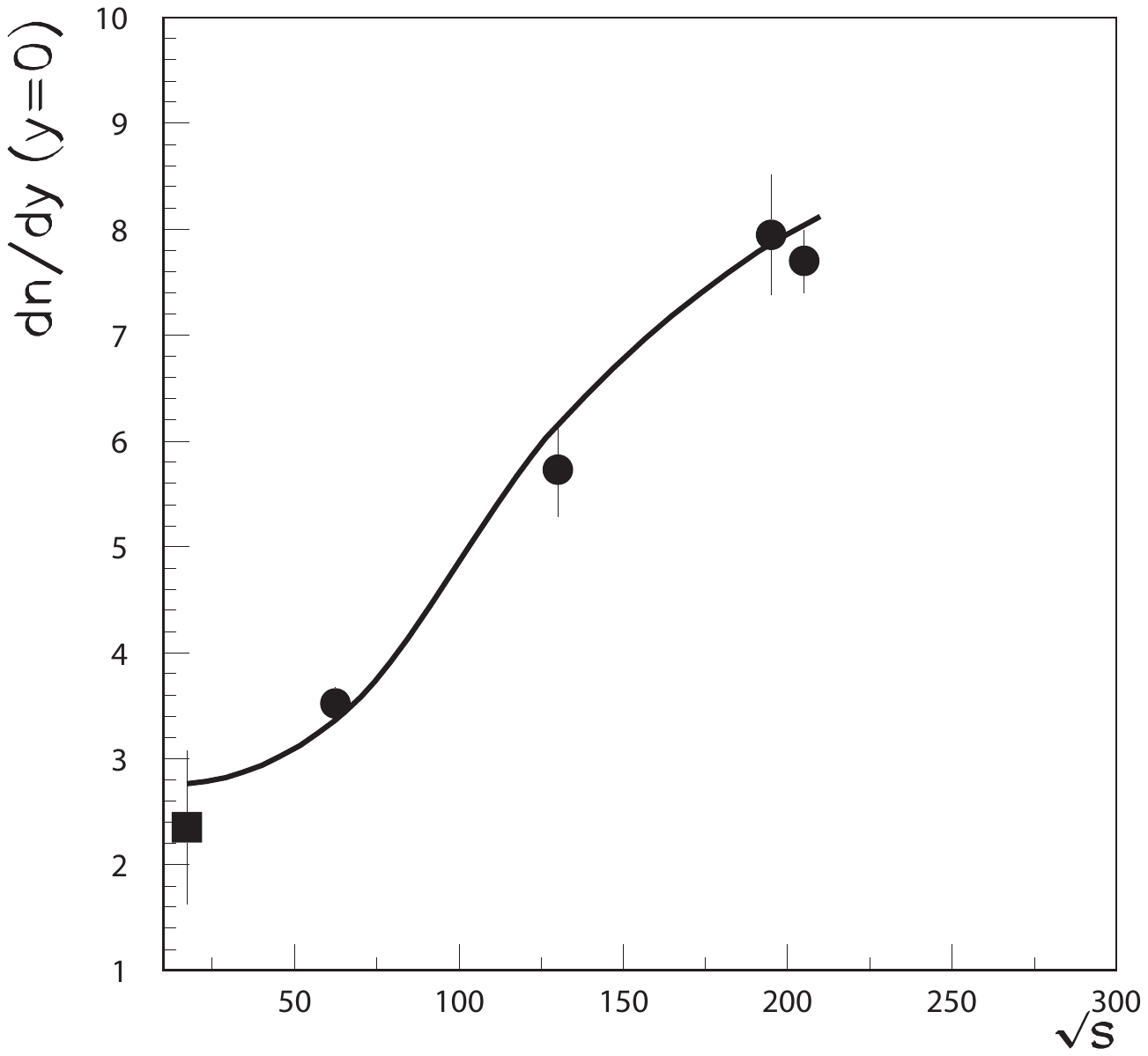}
\vskip -5.75cm
\caption{\label{fig:phi158ys}
Experimental data on the energy dependence of inclusive density for $\phi$-meson
production in the midrapidity region, $dn/dy$, $|y|\sim 0$, in $PbPb$ (closed box)~\cite{afanas, anticic}
and $AuAu$ (closed circles)~\cite{starphi, star2009, star130} collisions. The theoretical
curve represents the results of corresponding calculations for $AuAu$ collisions
based on the QGSM}
\end{figure}

Fig.~\ref{fig:phi158ys} shows experimental data on the energy dependence of the inclusive density
in the midrapidity region, $dn/dy\ (y=0)$,
in $PbPb$ (NA49 Collaboration, $\sqrt{s}= 17.3\ GeV$~\cite{afanas, anticic}, closed box)
and $AuAu$ collisions (STAR and PHENIX Collaborations, $\sqrt{s}= 62.4\ GeV$~\cite{star2009},
130~\cite{star130, star130adl}, and 200~$GeV$~\cite{starphi, star2009}, closed circles).
The theoretical curve has been calculated for $AuAu$ collisions by using the QGSM formalism.

A rather strong energy dependence of the inclusive density being considered stems from the fact that
the $\phi$-meson mass is quite sizable, and thus the minimal value of $x_{\pm}$ in Eq.~(\ref{eq:eq4})
decreases substantially as the initial energy grows. This leads to the corresponding broadening
of the integration domain in Eq.~(\ref{eq:eq5}) and, accordingly, to the growth of the inclusive density
in the midrapidity region.

\begin{figure}[htb]
\vskip -10.5cm
\hskip 2.5cm
\includegraphics[width=1.0\hsize]{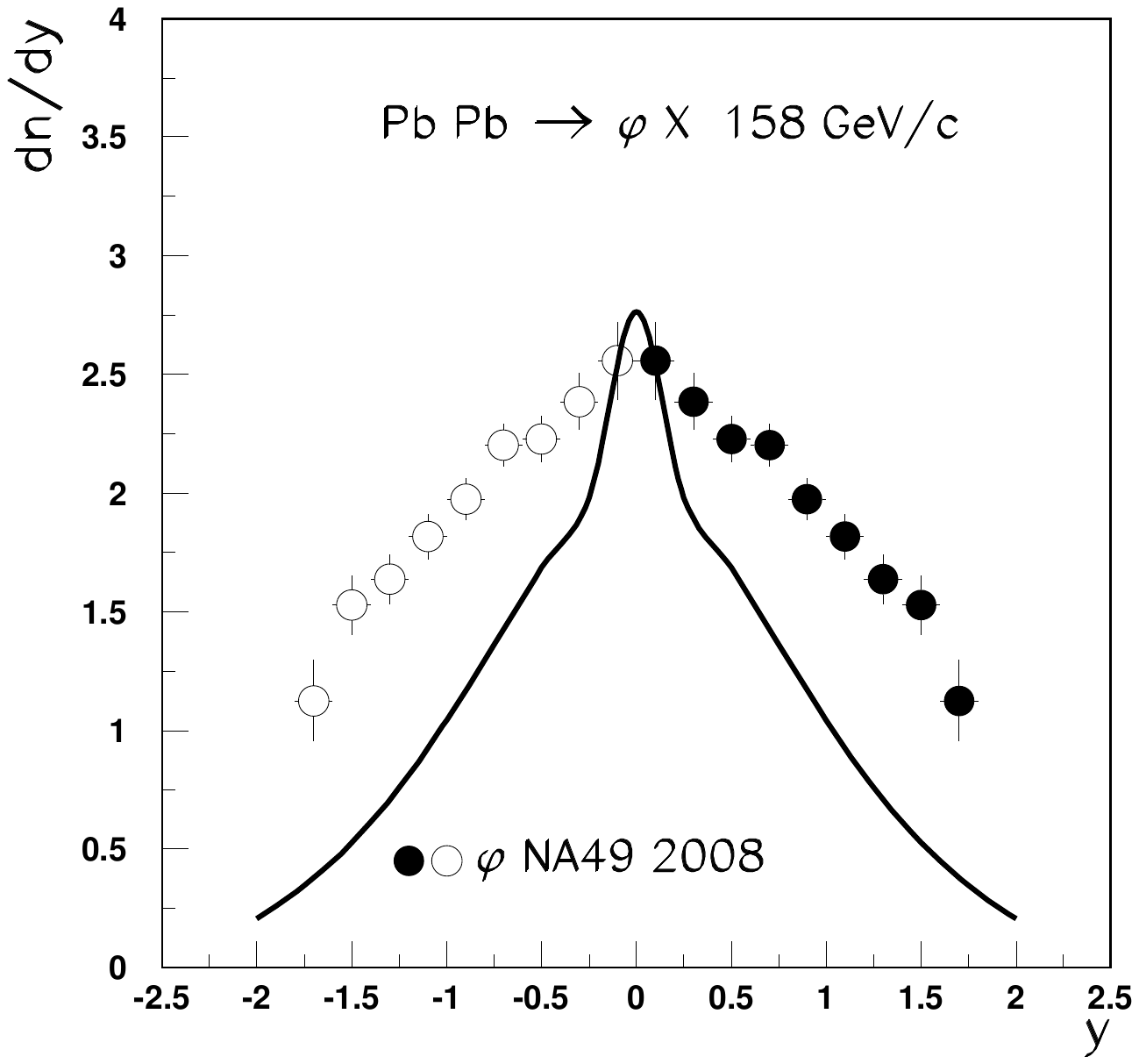}
\vskip -0.75cm
\caption{\label{fig:phi158y}
Experimental data on the rapidity dependence of the density for $\phi$-meson production, $dn/dy$,
in $PbPb$ collisions at the momentum of 158~GeV/c~\cite{afanas, anticic} (points) along
with the results of corresponding calculation based on the QGSM.}
\end{figure}

Experimental data on the rapidity spectrum, $dn/dy$,  of $\phi$-mesons produced
in $PbPb$ collisions at 158 GeV/c~\cite{afanas, anticic} are shown in Fig.~\ref{fig:phi158y}, 
along with the results of corresponding calculations based on the QGSM.

The agreement between the results of the theoretical calculations and experimental data
is quite resonable at $y=0$, but, as the rapidity $y$ grows, the theoretical curve goes down
more quickly than the experimental data. The reason for this possibly lies on the fact that at this 
energy it would be necessary to take into account the contribution of the quasi-two-particle channel,
which should be considered separately.

\section{The $\phi$-meson production on nuclear targets at LHC energies}
\label{sec:sec5}

In the previous section, it was shown that, in the case of $\phi$-meson
production, the inelastic-shadowing effects are very weak at RHIC energies,
being virtually invisible against the experimental errors. 

In this section we analyze the significance of the shadowing contribution for $\phi$-meson
production in the LHC energy range.

In ref.~\cite{CKTr}, it was explained that starting from RHIC energies,
the inclusive density for the production of secondary particles exhibits
significant saturation effects, both in $pPb$ and in $PbPb$ collisions,
what has been since experimentally confirmed~\cite{MPSd, Phob, Phen}.

Saturation effects can be explained by inelastic-shadowing corrections
connected to the multipomeron interactions~\cite{CKTr}, which
are negligible at low energies due to the suppression of the longitudional
part of the nuclear form factor.

In the case of interaction with nuclei, the mean number of Pomerons is large, and even
at the RHIC and LHC energies their interactions become significant. Since 
the growth of the initial energy weakens the suppression of the longitudinal part
of the nuclear factor, the inelastic-shadowing corrections become more and more
significant as the initial energy increases. 

The calculations of inclusive densities and multiplicities can be performed in the percolation approach
(with accounting for the inelastic shadowing) in both $pp$~\cite{CP1, CP2} and heavy ion
collisions~\cite{CP2, CP3}. The results of these calculations are in good agreement with the experimental 
data over a broad energy region.

The percolation approach assumes that two or several Pomerons overlap one another in the transverse space 
and merge together in a single Pomeron. As a result, and given a certain value of the transverse radius of 
the interaction region, when the density of Pomerons in that interaction region becomes large enough,
internal partons (quarks and gluons) from different nucleons can merge, leading to the saturation of the 
inclusive density of final-state particles. As the energy grows, this effect persists up to the overlap 
of all Pomerons~\cite{Dias, Paj, Braun}.

Technically, a more direct way to take into account the percolation effects in the QGSM~\cite{MPSd} 
is to consider the maximal number of cut Pomerons in the midrapidity region, $n_{max}$, that were emitted by 
one nucleon. If the number of emitted Pomerons is less than $n_{max}$, then different final states of produced 
particles arise upon cutting them. The contributions of all the diagrams with $n \leq n_{max}$ are then taken 
into account in just the same way as at low energies. According to the unitarity constraint, a larger number 
of Pomerons $n > n_{max}$ can be emitted. Because of final-state interaction (merger), however, the same final 
state as in the case of cutting $n_{max}$ Pomerons arises upon cutting these Pomerons at the quark-gluon stage.

With this prescription, the QGSM calculations become straightforward,
and their results are similar to those of calculations in the percolation model.
According to this scenario, the QGSM provides a reasonable agreement with
the experimental data on the inclusive spectra of light secondary particles at
a value $n_{max} = 13$ in the case of RHIC energies (see ref.~\cite{MPSd}),
and at $n_{max} = 21$ in the case of LHC energies~\cite{MPSpPb}.
These values of $n_{max}$ reflect the fact that the merger of Pomerons is quite efficient
in the case of $\pi^{\pm}$, $K^{\pm}$, $p$, and $\bar{p}$ production.
In the absence of inelastic shadowing, the maximal number of Pomerons is substantially larger,
which leads to higher values of inclusive densities~\cite{AMPSpl, MPSd}.

The number of strings that determines particle production grows with increasing initial energy
even if the percolation effects are included. This was explained in ref.~\cite{JDDCP}.
The point is that the number of Pomerons that could make a significant contribution
to the spectra in the absence of inelastic shadowing grows fast with increasing initial energy,
and an increase in the value of the parameter $n_{max}$ does not prevent the growth
of the relative contribution of inelastic shadowing.

Below we compare the experimental data on $\phi$-meson production in collisions on nuclear targets
at LHC energies, with the results of the corresponding QGSM calculations.

\begin{figure}[htb]
\vskip -10.25cm
\hskip 2.5cm
\includegraphics[width=1.0\hsize]{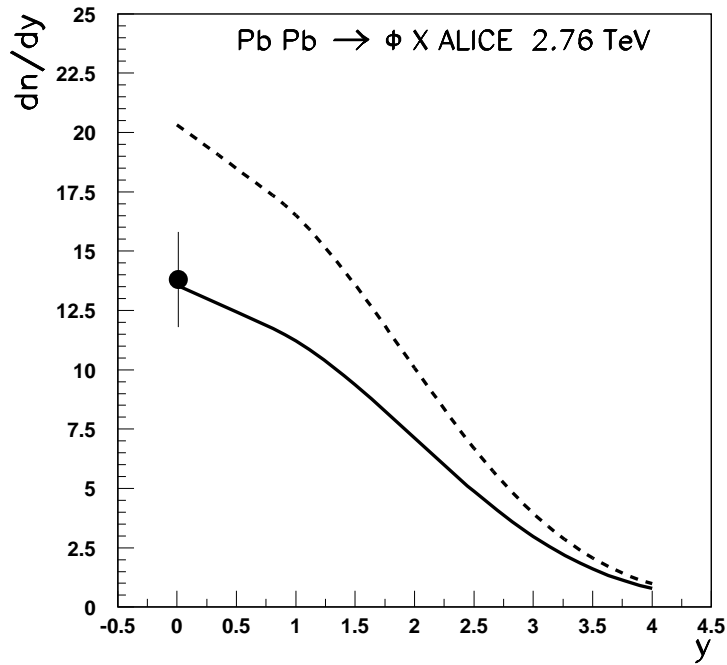}
\vskip -0.75cm
\caption{\label{fig:phi276TeV}
Comparison of the QGSM prediction for the rapidity dependence of the density of $\phi$-meson production,
$dn/dy$, in central $PbPb$ collisions at $\sqrt{s}$ = 2.76~Tev (solid curve) with one experimental point by
ALICE Collaboration~\cite{alicepb}. The dashed curve represents the result of the QGSM calculation without including the 
inelastic-shadowing effect.}
\end{figure}

In Fig.~\ref{fig:phi276TeV} we present the result of the QGSM calculations for the rapidity dependence of the 
density dn/dy of $\phi$-meson production in $PbPb$ collisions, for centrality 5\%  with (solid curve) 
and without (dashed curve) including the inelastic-shadowing effect. One can see that the
inelastic-shadowing contribution is sizable in the midrapidity region, but it decreases sharply as the rapidity $y$ grows.
Also in Fig.~\ref{fig:phi276TeV}, the QGSM prediction~\cite{AMShPhi2} is compared with the
experimental point on $dn/dy$ at $y=$0 for $\phi$-meson production at $\sqrt{s}$ = 2.76~TeV recently published by
the ALICE Collaboration~\cite{alicepb}.

In Table~4 we present the results of the QGSM calculations for LHC energies, obtained without using any additional
parameter with respect to the calculations at lower energies, so these results can be considered as theoretical
predictions. The comparison with the corresponding available experimental data is also shown. 
Unfortunately, only one experimental point by the ALICE Collaboration~\cite{alicepb} is currently available on the $\phi$-meson
production density $dn/dy$ ($|y|\leq $ 0.5) in $PbPb$ collisions at 
$\sqrt{s}$ = 2.76~TeV and centrality 5\%, and the same occurs for $\phi$-meson production
in $pPb$ collisions at $\sqrt{s}$ = 5.02~Tev for non-single diffraction (NSD) events~\cite{alippb5t}.

When comparing the experimental point by the ALICE Collaboration~\cite{alicepb} on the $\phi$-meson
production density $dn/dy$ ($|y|\leq $ 0.5) in $PbPb$ collisions at $\sqrt{s}$ = 2.76~Tev and centrality 5\%,
with the result of corresponding QGSM calculation, the value of $dn/dy$ ($|y|\leq 0.5$) = 13.57 is obtained,
which agrees with the experimental data, at $n_{max}$ = 37.
The calculation at an assimptotically large value of $n_{max}$ (that is, in the absence of inelastic
shadowing), yields a value of $dn/dy$ ($|y|\leq 0.5$) $\simeq$ 20.5. The inelastic-shadowing effect
reduces the value of $dn/dy$ ($|y|\leq 0.5$) by about 1.5. At this energy, the 
$\pi^{\pm}$, $K^{\pm}$, $p$, and $\bar{p}$ spectra decrease by a factor of about two because of
a larger contribution of inelastic shadowing.

On the other hand, the QGSM calculation for the density $dn/dy$ of $\phi$ mesons 
produced in $pPb$ collisions at $\sqrt{s}$ = 5.02~TeV, to compare with the 
corresponding experimental point by the ALICE Collaboration~\cite{alippb5t} at 
the same energy, for non-single diffraction (NSD) events, were made 
with $n_{max}$ = 32. The different values of $n_{max}$ are connected with different 
initial beams (the number of cutted Pomerons in $PbPb$ and $pPb$ collisions) and with
different initial energies.

Since once the value of $n_{max}$ is fixed, the secondary $\phi$-meson production in 
proton-nucleus and nucleus-nucleus collisions can be calculated without any additional 
parameter with respect to the corresponding calculations for $pp$ collisions.

As we can see in Table~4, QGSM gives a rather good description of these ALICE Collaboration points,
but, of course, new experimental data on $\phi$-meson production at LHC energies, both for $pPb$
and $PbPb$ collisions, will be crucial in order to draw final conclusions.
\vskip 0.5cm

\begin{table}
\caption{\label{Table4} Experimental point by the ALICE Collaboration~\cite{alicepb}
on $dn/dy$, $|y|\leq 0.5$, for $\phi$-meson production in $PbPb$ collisions at 
$\sqrt{s}$ = 2.76~TeV, compared to the result of corresponding QGSM calculation. 
We also compare the QGSM result with the experimental point on $pPb$ collisions at $\sqrt{s}$ = 5.02~TeV,
for non-single diffraction (NSD) events~\cite{alippb5t}.}
\vspace{-0.5cm}
	
\begin{center}
\begin{tabular}{lllll}
\hline
Reactions &Centrality & Energy & Experimental data & QGSM \\
	&  &  $\sqrt{s}$, $TeV$ & $dn/dy$ ($|y|\leq 0.5$)  &  \\
\hline
Pb + Pb & 0$-$5\% & 2.76 & 13.8 $\pm$ 0.5 $\pm$ 1.7 $\pm$ 0.1 \cite{alicepb} & 13.57  \\

p + Pb & NSD & 5.02 & 0.1344 $\pm$ 0.005 $\pm$ 0.0069 $\pm$ 0.0081 \cite{alippb5t}& 0.14 \\
\hline
\end{tabular}
\end{center}
\end{table}
\vspace {-1.cm}

\section{Conclusions}
\label{sec:Conclusions}

The QGSM provides a reasonable description of Feynman $x_F$ and rapidity $y$ spectra of 
$\phi$-meson production for the interaction of different hadron beams with a nucleon target
in a wide energy region, by using for the only unknown parameter in the QGSM analysis, the
normalization parameter $a_{\phi}$. The value $a_{\phi}$ = 0.11 is determined by comparison
with experimental data in the energy range up to RHIC, where the screenig contribution is negligible
for $\phi$-meson production, and then used for $\phi$-meson production processes at higher energies.

At LHC energies, a new parameter $n_{max}$, connected with the number of cut pomerons,
appears in the calculation of the inelastic nuclear shadowing contribution to $\phi$-meson production.

The QGSM prediction for the density $dn/dy$ at LHC energies in $pp$ collisions are compared with 
recent experimental data at $\sqrt{s}$ = 2.76 and 7~TeV in the midrapidity region.
The theoretical results based on the QGSM calculations for the $\phi/\pi^-$ and $\phi/K^-$
cross section ratios present a reasonable agreement with the corresponding experimental data
in a wide interval of the beam energy, going up to the LHC range.
Since the theoretical ratios $\phi/\pi^-$ and $\phi/K^-$ grow with energy from
$\sqrt{s}$= 0.9 to 7~Tev, while the experimental points do not show this growth,
and, since, in particular, the prediction based on the QGSM for the ratio $\phi/\pi^-$
at $\sqrt{s}$ = 0.9~TeV 
is in agreement with the experiment, some discrepancy of the QGSM calculations with the experimental
data appearing at the LHC energy of $\sqrt{s}$ = 7 TeV.

Concerning the interactions of protons and nuclei with nuclear targets, the QGSM provides
a reasonable description of $\phi$-meson production up to the RHIC energies without recourse to
any additional parameter, and without the inclusion of inelastic-shadowing effects. 

The effect of inelastic nuclear shadowing is markedly weaker in the case of $\phi$-meson
production than in the case of the production of other particles,
such as $\pi^{\pm}$, $K^{\pm}$, $p$, and $\bar{p}$, and becomes observable at higher energies
of$ \sqrt{s}\ \geq$ 0.9~Tev. This behaviour can be explained by the fact that the mass of the strange quark 
is not very small.
\vskip 0.75 cm

{\bf Acknowledgements}
This work has been supported by Russian RSCF grant No. 14-22-00281,
by Ministerio de Ciencia e Innovaci\'on of Spain under project
FPA2014-58293-C2-1-P, and Maria de Maeztu Unit of Excellence MDM-2016-0692,
and by Xunta de Galicia, Spain, under 2015-AEFIS (2015-PG034), AGRUP2015/11.

\setcounter{equation}{0}
\renewcommand{\theequation}{A.\arabic{equation}}

\vspace{-0.35cm}
\section*{Appendix A: Pomeron pole and multipomeron-exchange cross sections}
\label{sec:appendixA}
\vspace{-0.1cm}

The detailed description of the process of construction of the multipomeron-exchange amplitude 
for hadron-hadron scattering at high energies in the classical reggeon diagram technique 
is given in references~\cite{AMShPhi1, KTM, K20, KTMS, Volk}. 

In the case of a supercritical Pomeron with 
\begin{equation}
\alpha_P(t) = 1 + \Delta + \alpha'_P \, t\;,\,\, \Delta > 0\; ,
\label{eq:eq6}
\end{equation}
one obtains the correct asymptotic behaviour, $\sigma_{tot} \sim \ln^2s$.
The one-Pomeron contribution to $\sigma^{tot}_{hN}$ equals:
\begin{equation}
\sigma_P = 8\pi \gamma e^{\Delta \xi},\ \ \ \xi = \ln s/s_0 \;,
\label{eq:eq7}
\end{equation}
where $\gamma$ is the Pomeron coupling, and $\sigma_P$ rises with
energy as $s^{\Delta}$. To obey the $s$-channel unitarity, and the
Froissart bound in particular, this contribution should be screened by
the multipomeron discontinuities. A simple quasi-eikonal treatment
\cite{Kar3} yields:
\begin{equation}
\sigma^{tot}_{hN} = \sigma_P f(z/2) \;,\,\, \; \sigma^{el}_{hN} =
\frac{\sigma_P}{C} [f(z/2) - f(z)] \;,
\label{eq:eq8}
\end{equation}
where
\begin{eqnarray}
f(z) & = & \sum^{\infty}_{k=1} \frac1{k \cdot k !} (-z)^{k-1} =
	\frac1z \int ^z_0 \frac{dx}x (1-e^{-x}) \;, \label{eq:eq9}\\ z & = &
\frac{2C\gamma}{\lambda} e^{\Delta \xi} \;,\,\, \lambda = R^2 + \alpha'_P
\xi \;.
\label{eq:eq10}
\end{eqnarray}
Here, $R^2$ is the radius of the Pomeron, and $C$ is the quasi-eikonal
enhancement coefficient~(see~\cite{Kar1}).
At asymptotically high energies ($z \gg 1$), we obtain:
\begin{equation}
\sigma^{tot}_{hN} = \frac{8\pi \alpha'_P \Delta}{C} \xi^2 \;,
\sigma^{el}_{hN} = \frac{4\pi \alpha'_P \Delta}{C^2} \xi^2 \;,
\label{eq:eq11}
\end{equation}
according to the Froissart limit~\cite{Froi}.

The values of Pomeron parameters were fixed in ~\cite{AMShPhi1, Sh} on the base of 
a Regge fit~\cite{Volk} of high-energy hadron-nucleon scattering, by including into the
analysis the by then new data from colliders:
\begin{equation}
\Delta = 0.139 \;, \;\; \alpha'_P = 0.21\ {\rm GeV}^{-2} \;,
\label{eq:eq12}
\end{equation}
$$\gamma_{pp} = 1.77\ {\rm GeV}^{-2} \;, \;\;R^2_{pp} =  3.18\ {\rm GeV}^{-2}
\;, \;\;C_{pp} = 1.5,$$
$$\gamma_{\pi p} = 1.07\ {\rm GeV }^{-2} \;, \;\;R^2_{\pi p} =
2.48\ {\rm GeV}^{-2}
\;, \;\;C_{\pi p} = 1.65.$$
The error bars of the parameters are not presented, since they are strongly 
correlated.
With these values for the set of parameters in $pp$ collisions, one obtains a value of
$\sigma^{tot}_{pp}\simeq$ 94 mb at the LHC energy of $\sqrt{s}$ = 7 TeV,
that is only slightly smaller than the experimental value
$\sigma^{tot}_{pp}$ = 98.3 $\pm$ 2.8 mb~\cite{totem}.

\setcounter{equation}{0}
\renewcommand{\theequation}{B.\arabic{equation}}

\vspace{-0.35cm}
\section*{Appendix B: Quark and diquark distributions in hadrons in QGSM}
\label{sec:appendixB}
\vspace{-0.1cm}

The QGSM considers hadrons as consisting of constituent quarks, so we
cannot use hadron structure functions obtained from hard processes.
Usually, it is assumed in DTU that a proton consists
of a valence quark $q$ and a diquark $qq$. The diquark $qq$
contains not only two valence quarks, but also some part of gluon fileld, in what is called  
string junction~(SJ)~\cite{IOT, RV}. In this case, the diquark average momentum
is larger than twice the momentum of a valence quark. In the QGSM a
proton can also contain several sea quark-antiquark pairs.

\begin{figure}[htb]
\vskip -24.0cm
\hskip -3.5cm
\includegraphics[width=1.6\hsize]{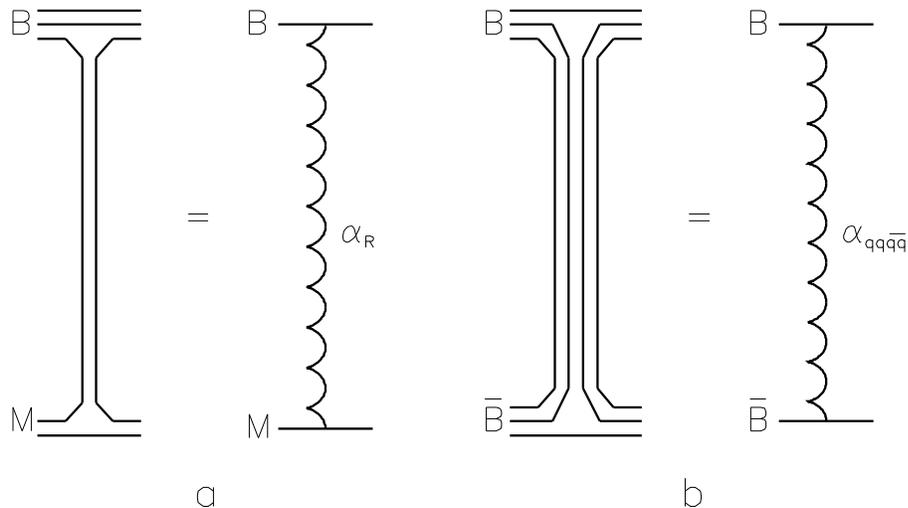}
\vskip -1.0cm
\caption{\label{fig:Planardiags}
Planar diagrams which determine the low-$x$ asymptotic behaviour ($a$) of
the valence quark distribution in a nucleon, and ($b$) of the diquark distribution
in a nucleon.}
\end{figure}

In QGSM the form of the functions $u(x,n)$ is determined by the
corresponding Regge asymptotic behaviours in the regions $x \to 0$ and
$x \to 1$~\cite{CST, Kai3}. As an example, let us consider the
diagram with annihilation of one quark from the fast nucleon
on a meson target, that is shown in Fig.~\ref{fig:Planardiags}a.

The contribution of this process to the total inelastic $hN$ cross
section is proportional to $e^{[\alpha_R(0) - 1]\Delta y}$, where
$\alpha_R(0) \simeq 0.5$ is the intercept parameter of the corresponding
non-vacuum Regge trajectory, and $\Delta y$ is the difference between the
rapidities of the colliding particles. On the other hand, and starting from the
parton model, we can consider this process from a different
point of view~\cite{Kai3}; {\it i.e.}, a slow valence quark exists in the incident fast nucleon
that is $\Delta y$ rapidity distant away
from the valence diquark, the probability of such a configuration
depending on $\Delta y$.
This slow valence quark annihilates with a
target antiquark with a rather large probability, what
corresponds to the cut of the diagram in Fig.~\ref{fig:Planardiags}a. Then,
the probability to find a slow valence quark in the fast nucleon
at a distance $\Delta y$, {\it i.e.} with $x\simeq e^{-\Delta y} \to 0$,
is equal to
\begin{equation}
x u_q(x) \sim x^{1 - \alpha_R(0)}\;,\; x \to 0 \;.
\label{eq:eqA1}
\end{equation}
The probability of finding a valence quark with $x \to 1$ in the nucleon
is determined by the probability of finding a slow diquark. This
can be detected in the process of $N\bar{N}$ annihilation that is
shown in Fig.~\ref{fig:Planardiags}b. The intercept of the corresponding Regge trajectory
$\alpha_{qq\bar{q}\bar{q}}$ can be calculated as~\cite{Kai, Kai3}
$\alpha_{qq\bar{q}\bar{q}}(0) = -\alpha_R(0) + 2 \alpha_B(0)$, where
$\alpha_B(0) \simeq -0.5$ is a parameter of the nucleon Regge trajectory.
Then,
\begin{equation}
x u_q(x) \sim (1-x)^{1 + \alpha_R(0) - 2\alpha_B(0)} \;,\; x \to 1 \;.
\label{eq:eqA2}
\end{equation}

The quark distribution in the intermediate $x$ region can be estimated
with the help of a simple interpolation,
\begin{equation}
u(x) = C x^{\alpha} (1-x)^{\beta},
\label{eq:eqA3}
\end{equation}
with both limits at $x \to 0$ and $x \to 1$ given by eqs.~\ref{eq:eqA1} and
\ref{eq:eqA2}. The normalization factor $C$ is determined by the condition
\begin{equation}
\int u_i(x,n) dx = 1 \;,
\label{eq:eqA4}
\end{equation}
leading to
\begin{equation}
C = \frac{\Gamma(\alpha + \beta +2)}{\Gamma(\alpha + 1)
\Gamma(\beta + 1)} \;.
\label{eq:eqA5}
\end{equation}

The numerical calculations account for the fact that the distribution
of valence $d$-quark in the proton is softer than the
distribution of valence $u$-quarks. This can be done by including
into $u_d(x,n)$ an additional factor $(1-x)$ with respect to
$u_u(x,n)$.
The diquark distributions can be derived from the quark
distributions by substituting $x \to (1-x)$.

The diquark and quark distribution functions depend on the number $n$ of cut Pomerons
in the diagrams being considered.

As a consequence of momentum conservation, for each value of $n$ one has
\begin{equation}
\sum_i \langle x_i \rangle = \sum_i \int u_i(x,n) xdx = 1 \;.
\label{eq:eqA6}
\end{equation}

In the case of $n > 1$, {\it i.e.}, for multipomeron exchange
the distributions of valence quarks and diquarks are softened with respect to 
the case of  $n > 1$, due to the appearance of a sea quark contribution. 
For arbitrary $n$, one has~\cite{KTMS}:
\begin{eqnarray}
u_{uu}(x,n) & = & C_{uu} x^{\alpha_R-2\alpha_B+1}
(1-x)^{-\alpha_R+\frac{4}{3}(n-1)}, \label{eq:eqA7}\\
u_{ud}(x,n) & = & C_{ud} x^{\alpha_R-2\alpha_B} (1-x)^{-\alpha_R+n-1},
\label{eq:eqA8}\\
u_u(x,n) & = & C_u x^{-\alpha_R} (1-x)^{\alpha_R-2\alpha_B+n-1},
\label{eq:eqA9}\\ 
u_d(x,n) & = & C_d x^{-\alpha_R} (1-x)^{\alpha_R-2\alpha_B+1+\frac{4}{3}(n-1)},
\label{eq:eqA10} \\
u_s(x,n) & = & C_d x^{-\alpha_R} (1-x)^{\alpha_R-2\alpha_B+n-1} \;.
\label{eq:eqA11}
\end{eqnarray}

The quark distributions in pion can be obtained in a similar way, e.g., 
for the case of $\pi^{-}$ mesons, we use the distributions~\cite{KTMS}:
\begin{eqnarray}
u_{d}(x,n) & = & u_{\bar{u}}(x,n) = C_{d} x^{-\alpha_{R}} (1 - x)^
{-\alpha_{R} + n - 1} \;, \label{eq:eqA12} \\
u_{u}(x,n) = u_{\bar{d}}(x,n) & = & C_{u} x^{-\alpha_{R}} (1 - x)^
{-\alpha_{R} + n - 1}[1 - \delta \sqrt{1 - x}] \;,\; n>1 \;, \label{eq:eqA13}\\
u_{\bar{s}}(x,n) & = & C_{\bar{s}} x^{-\alpha_{R}} (1 - x)^
{n - 1} \;, \;\;n>1 \;. \label{eq:eqA14}  
\end{eqnarray}

\setcounter{equation}{0}
\renewcommand{\theequation}{C.\arabic{equation}}

\vspace{-0.35cm}
\section*{Appendix C: Quark and diquark framentation functions in vector $\phi$-mesons in QGSM}
\label{sec:appendixB}
\vspace{-0.1cm}

First, it has to be noted that the fragmentation functions $G_i^h(z)$ used
in QGSM are related to the standard fragmentation functions $D(z)$
by the equation $G(z)=zD(z)$~\cite{K20, Kai}.

For the $\phi$-meson production we use the following quark fragmentation
functions~\cite{aryer}, that were obtained by using the Reggeon counting rules 
and the simplest extrapolation~\cite{Kai}:
\begin{equation}
G_{u}^{\phi} = G_{d}^{\phi} = G_{\bar u}^{\phi} = G_{\bar d}^{\phi} =
a_{\phi}\cdot(1-z)^{\lambda - \alpha_R - 2 \alpha_{\phi}+2}, \;\; \\ 
\label{eq:eq15}
\end{equation}
\begin{equation}
G_{s}^{\phi} = G_{\bar s}^{\phi} =  a_{\phi}\cdot(1-z)^{\lambda - \alpha_{\phi}}.
\label{eq:eq16}
\end{equation}
The diquark fragmentation functions into $\phi$-mesons have the form:
\begin{equation}
G_{uu}^{\phi} = G_{ud}^{\phi} = 
a_{\phi}\cdot(1-z)^{\lambda+\alpha_R -2(\alpha_R +\alpha_{\phi})} \ \ ,
\label{eq:eq17}
\end{equation}
where parameters $\alpha_R$ = 0.5 and $\alpha_{\phi}$= 0. are the intercepts of the $\rho$
and $\phi$ Regge trajectories, respectively.

The parameter $\lambda$ is $\lambda=2\alpha^{\prime}_R <p_T^2> $, where 
$\alpha^{\prime}_R$ is the slope of the vector-meson trajectory, and $<p_T^2>$ is the average square 
transverse momenta of the produced mesons.

The only unknown value in the model is that of parameter $a_{\phi}$,
which corresponds the normalisation of the $\phi$ density in the central region of rapidity $y$.
This value is universal in the sense that it does not depend, neither of the 
energy, nor of the beam and target types of the collision processes.
The value of $a_{\phi}$ has been determined by comparing the results of the model 
calculations with the available experimental data 
on $\phi$ production from pion and proton beams (see section~\ref{sec:sec3}).
Thus, in the calculations we present here the value of $a_{\phi}=0.11$, which gives a good description
of most of those experimental data, has been used.

\end{document}